\documentclass{ws-ijbc}
\usepackage{ws-rotating}     
\usepackage{amsmath}
\usepackage{amssymb}
\usepackage{amsfonts}
\usepackage{graphicx}
\usepackage{graphics}
\usepackage{theorem}
\usepackage{color}
\usepackage{setspace}
\usepackage{makeidx}
\usepackage{mathrsfs}
\usepackage{amscd}
\usepackage{pstricks}
\usepackage{epsfig}
\usepackage{rotate}

\begin{document}

\catchline{}{}{}{}{} 

\markboth{Manos et al.}{Probing the local dynamics of periodic orbits}

\title{PROBING THE LOCAL DYNAMICS OF PERIODIC ORBITS BY THE
  GENERALIZED ALIGNMENT INDEX (GALI) METHOD}

\author{T.~MANOS} \address{University of Nova Gorica, School of Applied
Sciences,\\ Vipavska 11c, SI-5270, Ajdov\v{s}\v{c}ina, Slovenia\\ and\\ Center
for Research and Applications of Nonlinear Systems,\\ University of Patras,
GR-26500, Patras, Greece\\ thanosm@master.math.upatras.gr}

\author{CH.~SKOKOS} \address{Max Planck Institute for the Physics of Complex
Systems,\\ N\"{o}thnitzer Str. 38, D-01187, Dresden, Germany\\ and\\ Center
for Research and Applications of Nonlinear Systems,\\ University of Patras,
GR-26500, Patras, Greece\\ hskokos@pks.mpg.de}

\author{CH.~ANTONOPOULOS} \address{Interdisciplinary Center for Nonlinear
Phenomena and Complex Systems (CENOLI),\\ Service de Physique des Syst\`{e}mes
Complexes et M\'{e}canique Statistique,\\ Universit\'{e} Libre de Bruxelles,
1050, Brussels, Belgium\\cantonop@ulb.ac.be}

\maketitle


\begin{abstract}
  As originally formulated, the Generalized Alignment Index (GALI)
  method of chaos detection has so far been applied to distinguish
  quasiperiodic from chaotic motion in conservative nonlinear
  dynamical systems. In this paper we extend its realm of
  applicability by using it to investigate the local dynamics of
  \textit{periodic} orbits. We show theoretically and verify
  numerically that for \textit{stable} periodic orbits the GALIs tend
  to zero following particular power laws for Hamiltonian flows, while
  they fluctuate around non-zero values for symplectic maps. By
  comparison, the GALIs of \textit{unstable} periodic orbits tend
  exponentially to zero, both for flows and maps. We also apply the
  GALIs for investigating the dynamics in the neighborhood of periodic
  orbits, and show that for chaotic solutions influenced by the
  homoclinic tangle of unstable periodic orbits, the GALIs can exhibit
  a remarkable oscillatory behavior during which their amplitudes
  change by many orders of magnitude. Finally, we use the GALI method
  to elucidate further the connection between the dynamics of
  Hamiltonian flows and symplectic maps. In particular, we show that,
  using for the computation of GALIs the components of deviation
  vectors orthogonal to the direction of motion, the indices of stable
  periodic orbits behave for flows as they do for maps.
\end{abstract}

\keywords{GALI method, Periodic orbits, Hamiltonian flows, Symplectic maps}

\section{Introduction}\label{intro}

The method of the Generalized Alignment Indices (GALIs) was originally
introduced in \cite{GALI:2007} as an efficient chaos detection
method. To date, the GALI method has been successfully applied to a
wide variety of conservative dynamical systems for the discrimination
between regular and chaotic motion, as well as for the detection of
regular motion on low dimensional tori
\cite{LDI,ChrisBou,SkoBouAnto,MSB08,MSB09,BouManChris,SG10,MR10,MA11,GES11}.

In the present paper we extend and complete the study of the GALI
method, focusing on its behavior for the special case of periodic
orbits and their neighborhood in conservative dynamical systems.  The
detection of periodic orbits and the determination of their stability
are fundamental approaches for the study of nonlinear dynamical
systems, since they provide valuable information for the structure of
their phase space. In particular, stable periodic orbits are
associated with regular motion, since they are surrounded by tori of
quasiperiodic motion, while in the vicinity of unstable periodic
orbits chaotic motion occurs.

The GALI method is related to the evolution of several deviation
vectors from the studied orbit, and therefore is influenced by the
characteristics of the system's tangent space.  The main goal of the
paper is to determine the usefulness of the method for probing the
local dynamics of periodic orbits with different stability types.  We
manage to achieve this goal by deriving theoretical predictions for
the behavior of GALIs for stable and unstable periodic orbits. We also
verify numerically the validity of these predictions, by studying the
evolution of GALIs for periodic orbits of several Hamiltonian flows
and symplectic maps, clarifying also the connections of such dynamical
systems. In addition, we show how the properties of the index can be
used to locate stable periodic orbits, and to understand the dynamics
in the vicinity of unstable ones.

The paper is organized as follows: in the first two introductory
sections we recall the definition of the GALI, describing also its
behavior for regular and chaotic orbits (Sect.~\ref{sec:GALI}), and
report the several stability types of periodic orbits in conservative
systems (Sect.~\ref{sec:stability}). In Sect.~\ref{sec:G_PO} we first
study theoretically the behavior of the index for stable and unstable
orbits, and then present applications of the GALI to particular orbits
of Hamiltonian flows and symplectic maps. Sect.~\ref{sec:near} is
devoted to the dynamics in the neighborhood of periodic orbits, while
Sect.~\ref{sect:perp_dev_vec} is dedicated to the relation between
the GALIs of stable periodic orbits for flows and maps. Finally, in
Sect.~\ref{sec:conclusions}, we summarize our results.

\section{The Generalized Alignment Index (GALI)}
\label{sec:GALI}

Let us briefly recall the definition of the GALIs and their behavior
for regular and chaotic motion in conservative dynamical
systems. Consider an autonomous Hamiltonian system of $N$ degrees of
freedom ($N$dof), described by the Hamiltonian $H(q_1,q_2, \ldots,
q_N,p_1,p_2, \ldots, p_N)$, where $q_i$ and $p_i$, $i=1,2,\ldots,N$
are the generalized coordinates and conjugate momenta respectively. An
orbit in the $2N$-dimensional phase space $\mathcal{S}$ of this system
is defined by a vector $\vec{x}(t)=(q_1(t),q_2(t), \ldots,
q_N(t),p_1(t),p_2(t), \ldots, p_N(t))$, with $x_i=q_i$, $x_{i+N}=p_i$,
$i=1,2,\ldots,N$. The time evolution of this orbit is governed by
Hamilton's equations of motion
\begin{equation}
\frac{d \vec{x}}{dt}= \vec{\mathcal{V}}(\vec{x})= \left(
\frac{\partial H}{\partial \vec{p}}\, ,  - \frac{\partial
H}{\partial \vec{q}} \right),
\label{eq:Hameq}
\end{equation}
while the time evolution of an initial deviation vector
$\vec{w}(0)=(dx_1(0),\ldots, dx_{2N}(0))$ from the $\vec{x}(t)$
solution of Eqs.~(\ref{eq:Hameq}), obeys the variational equations
\begin{equation}
\frac{d \vec{w}}{dt} = \textbf{M}(\vec{x}(t)) \cdot\vec{w} \, ,
\label{eq:var}
\end{equation}
where $\textbf{M}= \partial \vec{\mathcal{V}} / \partial \vec{x}$ is
the Jacobian matrix of $\vec{\mathcal{V}}$.

Let us also consider a discrete time $t=n \in \mathbb{N}$ conservative
dynamical system defined by a $2N$-dimensional ($2N$D) symplectic map
$F$. The evolution of an orbit in the $2N$-dimensional space
$\mathcal{S}$ of the map is governed by the difference equation
\begin{equation}
\vec{x}(n+1)\equiv\vec{x}_{n+1}=F(\vec{x}_n).
\label{eq:map_gen}
\end{equation}
In this case, the evolution of a deviation vector $\vec{w}(n)\equiv
\vec{w}_n$, with respect to a reference orbit $\vec{x}_n$, is given by
the corresponding \textit{tangent map}
\begin{equation}
  \vec{w}(n+1)\equiv \vec{w}_{n+1}=
  \frac{\partial F}{\partial\vec{x}} (\vec{x}_n)\cdot\vec{w}_n.
\label{eq:w_map}
\end{equation}

For $N$dof Hamiltonian flows and $2N$D maps the Generalized Alignment
Index of order $k$ (GALI$_k$), $2 \leq k \leq 2N$, is determined
through the evolution of $k$ initially linearly independent deviation
vectors $\vec{w}_k(0)$. To avoid overflow problems, the resulting
deviation vectors $\vec{w}_k(t)$ are continually normalized, but their
directions are kept intact. Then, according to \cite{GALI:2007}
GALI$_k$ is defined as the volume of the $k$-parallelogram having as
edges the $k$ unit deviation vectors $\hat{w}_i(t)=\vec{w}_i(t)/
\|\vec{w}_i(t) \|$, $i=1,2,\ldots,k$, determined through the wedge
product of these vectors as
\begin{equation}
\mbox{GALI}_k(t)=\| \hat{w}_1(t)\wedge \hat{w}_2(t)\wedge \cdots
\wedge\hat{w}_k(t) \|,
\label{eq:GALI}
\end{equation}
with $\| \cdot \|$ denoting the usual norm. From this definition it is
evident that if at least two of the deviation vectors become linearly
dependent, the wedge product in Eq.~(\ref{eq:GALI}) becomes zero and
the GALI$_k$ vanishes.

In the $2N$-dimensional phase space $\mathcal{S}$ of an $N$dof
Hamiltonian flow or a $2N$D map, regular orbits lie on $s$-dimensional
tori, with $2\leq s \leq N$ for Hamiltonian flows, and $1\leq s \leq
N$ for maps.  For such orbits, all deviation vectors tend to fall on
the $s$-dimensional tangent space of the torus on which the motion
lies. Thus, if we start with $k\leq s$ general deviation vectors,
these will remain linearly independent on the $s$-dimensional tangent
space of the torus, since there is no particular reason for them to
become linearly dependent. As a consequence GALI$_k$ remains
practically constant and different from zero for $k\leq s$. On the
other hand, GALI$_k$ tends to zero for $k>s$, since some deviation
vectors will eventually have to become linearly dependent. In
particular, the generic behavior of GALI$_k$ for regular orbits lying
on $s$-dimensional tori is given by \cite{ChrisBou,SkoBouAnto}
\begin{equation}
  \mbox{GALI}_k (t) \propto \left\{ \begin{array}{ll} \mbox{constant} & \mbox{if
        $2\leq k \leq s$} \\ \frac{1}{t^{k-s}} & \mbox{if $s< k \leq 2N-s$} \\
      \frac{1}{t^{2(k-N)}} & \mbox{if $2N-s< k \leq 2N$} \\
\end{array}\right. .
\label{eq:GALI_order_all}
\end{equation}
Note that these estimations are valid only when the conditions stated
above are exactly satisfied. For example, in the case of 2D maps,
where the only possible torus is an $s=1$-dimensional invariant curve,
the tangent space is $1$-dimensional. Thus, the behavior of GALI$_2$
(which is the only possible index in this case) is given by the third
branch of Eq.~(\ref{eq:GALI_order_all}), i.e.~GALI$_2 \propto 1/t^2$,
since the first two cases of Eq.~(\ref{eq:GALI_order_all}) are not
applicable. From Eq.~(\ref{eq:GALI_order_all}) we deduce that, for the
usual case of regular orbits lying on an $N$-dimensional torus, the
behavior of GALI$_k$ is given by
\begin{equation}
  \mbox{GALI}_k (t) \propto \left\{ \begin{array}{ll} \mbox{constant} & \mbox{if
        $2\leq k \leq N$} \\ \frac{1}{t^{2(k-N)}} & \mbox{if $N< k \leq 2N$} \\
\end{array}\right. .
\label{eq:GALI_order_all_N}
\end{equation}

On the other hand, for a chaotic orbit all deviation vectors tend to
become linearly \textit{dependent}, aligning themselves in the
direction defined by the maximum Lyapunov characteristic exponent
(mLCE) and hence, in that case, GALI$_{k}$ tends to zero
\textit{exponentially} following the law \cite{GALI:2007}
\begin{equation}
\mbox{GALI}_k(t) \propto e^{-\left[ (\sigma_1-\sigma_2) + (\sigma_1-\sigma_3)+
\cdots+ (\sigma_1-\sigma_k)\right]t},
\label{eq:GALI_chaos}
\end{equation}
where $\sigma_1, \ldots, \sigma_k$ are the first $k$ largest Lyapunov
characteristic exponents (LCEs) of the orbit.

The GALI is a generalization of a similar indicator called the Smaller
Alignment Index (SALI) \cite{S01b,SkoAntoBouVrah,SABV04}, which has
been used successfully for the detection of chaos in several dynamical
systems
\cite{SESS04,PBS04,BS06,CLMV07,MSAB08,MSCHJD07,SHC09,MDSC10}. The
generalization consists in the fact that the GALIs use information of
\textit{more than two} deviation vectors from the reference orbit,
leading to a faster and clearer distinction between regular and
chaotic motion compared with SALI. In practice, SALI is equivalent to
GALI$_2$ since $\mbox{GALI}_2 \propto \mbox{SALI}$ (see Appendix B of
\cite{GALI:2007} for more details).

For the numerical computation of GALIs we consider the $k \times 2N $
matrix $\textbf{W}(t)$ having as rows the coordinates $w_{ij}(t)$ of
the unit deviation vectors $\hat{w}_i(t)$, $i=1,2,\ldots,k$,
$j=1,2,\ldots,2N$, with respect to the usual orthonormal basis
$\hat{e}_1= (1,0,0,\ldots,0)$, $\hat{e}_2= (0,1,0,\ldots,0)$,...,
$\hat{e}_{2N}= (0,0,0,\ldots,1)$ of the $2N$-dimensional phase space
$\mathcal{S}$. Thus, GALI$_k(t)$ can be evaluated as the square root
of the sum of the squares of the determinants of all possible $k
\times k$ submatrices of $\textbf{W}$ \cite{GALI:2007}
\begin{equation}
\mbox{GALI}_k=
\left\{\sum_{1 \leq i_1 < i_2 < \cdots < i_k \leq 2N} \left|
\begin{array}{cccc}
w_{1 i_1} & w_{1 i_2} & \cdots & w_{1 i_k} \\
w_{2 i_1} & w_{2 i_2} & \cdots & w_{2 i_k} \\
\vdots & \vdots &  & \vdots \\
w_{k i_1} & w_{k i_2} & \cdots & w_{k i_k} \end{array} \right|^2
\right\}^{1/2} . \label{eq:norm}
\end{equation}
Here the sum is performed over all possible combinations of $k$
indices out of $2N$, $|\cdot |$ denotes the determinant, and the
explicit dependence of all quantities on the time $t$ is omitted for
simplicity.

Equation (\ref{eq:norm}) is ideal for the theoretical determination of
the asymptotic behavior of GALIs for chaotic and regular orbits. It
has been used in \cite{ChrisBou,GALI:2007,SkoBouAnto} for the
derivation of Eqs.~(\ref{eq:GALI_order_all}) and
(\ref{eq:GALI_chaos}), and will be applied later on in
Sect.~\ref{sec:theory} for the determination of GALIs' behavior for
periodic orbits. However, from a practical point of view the
application of Eq.~(\ref{eq:norm}) for the numerical evaluation of
GALI$_k$ is not very efficient as it might require the computation of
a large number of determinants. In \cite{LDI,SkoBouAnto} a more
efficient numerical technique for the computation of GALI$_k$, which
is based on the Singular Value Decomposition of matrix $\textbf{W}(t)$
was presented. In particular, it has been shown that GALI$_k$ is equal
to the product of the singular values $z_i \geq 0$, $i=1,2,\ldots,k$
of $\textbf{W}^{\mathrm{T}}(t)$
\begin{equation}
\mbox{GALI}_k(t) = \prod_{i=1}^k z_i(t),
\label{eq:gali_svd}
\end{equation}
where $(^{\mathrm{T}})$ denotes the transpose matrix.

\section{Stability of Periodic Orbits}
\label{sec:stability}

Now, consider a $T$-periodic orbit (i.e.~an orbit satisfying
$\vec{x}(t+T)=\vec{x}(t)$) of an $N$dof Hamiltonian flow or of a $2N$D
symplectic map. Its linear stability is determined by the eigenvalues
of the so-called monodromy matrix $\textbf{Y}(T)$, which is obtained
from the solution of the variational equations for one period $T$ (see
for example
\cite{B69}\cite[Sect.~3.3]{LichtenbergL_1992}\cite{skokos:2001,H06}\cite[Chapt.~4,
5]{ChaosBook_08}). The monodromy matrix is
symplectic\footnote{$\textbf{Y}(T)$ satisfies the condition $
  \textbf{Y}(T)^{\mathrm{T}}\cdot \textbf{J}_{2N} \cdot \textbf{Y}(T)
  = \textbf{J}_{2N} $, with $ \textbf{J}_{2N}=
  \left[ \begin{array}{cc} \textbf{0}_{N} & \textbf{I}_{N} \\
      -\textbf{I}_{N} & \textbf{0}_{N}
\end{array}
\right]$, where $\textbf{I}_{N}$ is the $N\times N$ identity matrix
and $\textbf{0}_{N}$ is the $N\times N$ zero matrix.}, and its columns
correspond to linearly independent solutions of the equations that
govern the evolution of deviation vectors. In particular, the
evolution of an initial deviation $\vec{w}(0)$ from a $T$-periodic
orbit is given by
\begin{equation}
\vec{w}(iT)=\left[ \textbf{Y}(T) \right]^i \cdot \vec{w}(0), \,\,\,\,\,\,
i=1,2,\ldots\,\,.
\label{eq:w_ham_T}
\end{equation}

Due to the symplectic nature of the monodromy matrix and the fact that
its elements are real, the eigenvalues of $\textbf{Y}(T)$ have the
following property: if $\lambda$ is an eigenvalue then $1/\lambda$ and
the complex conjugate $\lambda^*$ are also eigenvalues. This property
shows that the eigenvalues $\lambda = 1$ and $\lambda = -1$ come in
pairs and that complex eigenvalues with modulus not equal to 1 always
appear in quartets.  When all eigenvalues are on the unit circle the
corresponding periodic orbit is said to be \textit{stable}. If there
exist eigenvalues off the unit circle the periodic orbit is
\textit{unstable}.

A few remarks on the connection of Hamiltonian systems with symplectic
maps are necessary at this point. Since autonomous Hamiltonian systems
are conservative, the constancy of the Hamiltonian function introduces
a constraint which fixes an eigenvalue of the monodromy matrix to be
equal to 1 and so, by the symplectic property, there must be a second
eigenvalue equal to 1. Thus, for an $N$dof Hamiltonian system there
are only $2(N-1)$ a priori unknown eigenvalues, and so we can reduce
our study to a $2(N-1)$-dimensional subspace of phase space
$\mathcal{S}$. This subspace is obtained by the well-known method of
the Poincar\'e surface of section (PSS)
(e.g.~\cite[p.~17-20]{LichtenbergL_1992}\cite[Sect.~3.1,
3.2]{ChaosBook_08}). The corresponding monodromy matrix of the
periodic orbit is also symplectic. Thus, in this sense, an $N$dof
Hamiltonian system is dynamically equivalent to $2(N-1)$D symplectic
map.

The different stability types of a periodic orbit in Hamiltonian
systems of 2dof and 3dof (or equivalently in 2D and 4D maps) have been
studied in detail in \cite{B69,H75,DM98,H06}, while the stability of
periodic orbits in higher dimensional conservative systems was
considered in \cite{HM87, HD98, skokos:2001}. Following the
terminology introduced in \cite{skokos:2001}, the general stability
type of a periodic orbit of an $N$dof Hamiltonian system, or
equivalently a $2(N-1)$D map, is denoted by
\begin{equation}
S_p U_m \Delta_l, \,\,\,\,\,\, \mbox{with} \,\,\,
p+m+2l=N-1,
\label{eq:sta_type}
\end{equation}
which means that $p$ couples of eigenvalues are on the unit circle,
$m$ couples are on the real axis and $l$ quartets are on the complex
plane but off the unit circle and the real axis. We conclude that a
periodic orbit is stable only when its stability type is $S_{N-1}$. In
all other cases the orbit is unstable since there exist eigenvalues of
the monodromy matrix off the unit circle. For example, in the case of
a 3dof Hamiltonian system or a 4D map a periodic orbit can be linearly
stable ($S_2$) or have three different types of instability: $S_1U_1$,
$U_2$, $\Delta_1$ (often called simple unstable, double unstable and
complex unstable respectively, see e.g.~\cite{ConMag:1985}).

\section{The Behavior of the GALI for Periodic Orbits}
\label{sec:G_PO}

\subsection{Theoretical treatment}
\label{sec:theory}

Let $\lambda_i$, $i=1,2,\ldots, 2N$ be the (possibly complex)
eigenvalues of the monodromy matrix $\textbf{Y}(T)$ of a $T$-periodic
orbit, ordered as $|\lambda_1| \geq |\lambda_2| \geq \cdots \geq
|\lambda_{2N}|$. Then, the corresponding LCEs $\sigma_i$,
$i=1,2,\ldots, 2N$ are given by \cite{BG_79,BFS_79,GALI:2007,Sko:LE}
\begin{equation}
\sigma_i = \frac{1}{T} \ln |\lambda_i|.
\label{eq:LCE_PO}
\end{equation}

In the case of unstable periodic orbits, where at least $| \lambda_1
|>1$, we get $\sigma_1>0$, which implies that nearby orbits diverge
exponentially from the periodic trajectory. Unstable periodic orbits
of non-integrable Hamiltonian systems and symplectic maps are located
inside chaotic domains. All non-periodic chaotic orbits in these
domains have the same spectrum of LCEs, which in general differs from
the spectrum of LCEs of the unstable periodic orbits of these domains.

For determining the behavior of GALI$_k$ for unstable periodic orbits,
one can apply the analysis presented in \cite{GALI:2007} for chaotic
orbits which also have $\sigma_1>0$. This approach leads to the
conclusion that GALI$_k$ of unstable periodic orbits tends to zero
exponentially following the law (\ref{eq:GALI_chaos})
\begin{equation}
  \mbox{GALI}_k(t) \propto e^{-\left[ (\sigma_1-\sigma_2) + (\sigma_1-\sigma_3)+
      \cdots+ (\sigma_1-\sigma_k)\right]t}.
\label{eq:GALI_chaos_upo}
\end{equation}

However, the case of stable periodic orbits needs a more careful
investigation. For this purpose let us consider an $N$dof Hamiltonian
system expressed in action-angle variables $J_i$, $\theta_i$,
$i=1,2,\ldots,N$. The equations of motion of a periodic orbit of this
system are
\begin{equation}
  \dot{J}_i  = -\frac{\partial H}{\partial \theta_i}= 0, \,\,\, \dot{\theta}_i = \frac{\partial H}{\partial J_i} = \omega_i(J_1,J_2,\dots, J_N),\,\,\, 1 \leq i \leq N.
\label{eq:a-a}
\end{equation}
The frequencies $\omega_i$ satisfy a relation of the form
\begin{equation}
\frac{\omega_1}{k_1}=\frac{\omega_2}{k_2}=\ldots=\frac{\omega_N}{k_N}=\Omega(J_1,J_2,\dots, J_N),
\label{eq:omega_per}
\end{equation}
where $k_i$, $i=1,2,\ldots,N$, are integer numbers and
$\Omega(J_1,J_2,\dots, J_N)=2\pi/T$ with $T$ being the period of the
orbit.  Equations (\ref{eq:a-a}) can be easily integrated to give
\begin{equation}
J_i(t)  =  J_{i0}, \,\,\, \theta_i(t)  =  \theta_{i0}+ \Omega(J_{10},J_{20},\dots, J_{N0}) k_i t, \,\,\, 1 \leq i \leq N,
\label{eq:a-a-sol}
\end{equation}
where $J_{i0}$, $\theta_{i0}$, $i=1,2,\ldots,N$ are the initial
conditions.

Let us now denote by $\xi_i$, $\eta_i$, $i=1,2,\ldots,N$, small
deviations from $J_i$ and $\theta_i$ respectively. Inserting
Eqs.~(\ref{eq:a-a}) and (\ref{eq:omega_per}) into the variational
equations of the Hamiltonian system we get
\begin{equation}
\dot{\xi}_i  =  0, \,\,\, \dot{\eta}_i   = k_i \sum_{j=1}^N  \Omega_{j} \xi_j, \,\,\, 1 \leq i \leq N,
\label{eq:twist_map_var}
\end{equation}
where $\Omega_{j}=\partial \Omega/ \partial J_j$ are computed for the
initial constant values $J_{j0}$, $j=1,2,\ldots, N$. Using as basis of
the $2N$-dimensional tangent space of the Hamiltonian flow the $2N$
unit vectors $\{\hat{v}_1,\hat{v}_2,\ldots,\hat{v}_{2N}\}$, such that
the first $N$ of them correspond to the $N$ action variables and the
remaining ones to the $N$ conjugate angle variables, any initial
deviation vector $\vec{w}_i(0)=(\xi_1^i(0),\xi_2^i(0),\ldots,
\xi_N^i(0),\eta_1^i(0),\eta_2^i(0),\ldots, \eta_N^i(0))$, evolves in
time as
\begin{equation}
  \vec{w}_i(t) = \sum_{j=1}^{N} \xi_j^i(0) \, \hat{v}_j +  \sum_{j=1}^{N} \left[ \eta_j^i(0) + \left( \sum_{l=1}^N  \Omega_{l} \xi_l^i (0) \right) k_j t \right] \,
\hat{v}_{N+j}.
\label{eq:order_dev_vec}
\end{equation}
From the above it readily follows that for sufficiently long times
$\|\vec{w}_i(t)\|\propto t$.

Let us now consider $k$, initially linearly independent, randomly
chosen, unit deviation vectors $\{\hat{w}_1,\ldots,\hat{w}_{k}\}$,
with $2\leq k \leq 2N$, and let $\textbf{W}$ be the matrix having as
rows the coordinates of these vectors with respect to the
$\{\hat{v}_1,\hat{v}_2,\ldots,\hat{v}_{2N}\}$ basis.  Defining by
$\mbox{\boldmath $\xi$}_i^{k}$ and $\mbox{\boldmath $\eta$}_i^{k}$,
$i=1,2,\ldots, N$, the $k\times 1$ column matrices
\begin{equation}
\mbox{\boldmath $\xi$}_i^{k} =
\left[ \begin{array}{cccc} \xi_i^1(0) &\xi_i^2(0) &\ldots
&\xi_i^k(0) \end{array}\right]^{\mathrm{T}}
 , \,\,\,
\mbox{\boldmath $\eta$}_i^{k} =
\left[ \begin{array}{cccc} \eta_i^1(0) &\eta_i^2(0) &\ldots
&\eta_i^k(0) \end{array}\right]^{\mathrm{T}}, \label{eq:xi_eta}
\end{equation}
$\textbf{W}(t)$ assumes the form
\begin{equation}
\textbf{W}(t) \propto \frac{1}{t^{k}} \cdot \textbf{W}^{k}(t)= \frac{1}{t^{k}} \left[
\begin{array}{cccccc}
\mbox{\boldmath $\xi$}_1^{k} & \ldots &
\mbox{\boldmath $\xi$}_N^{k} & \left( \mbox{\boldmath $\eta$}_1^{k} +
\left[ \displaystyle\sum_{l=1}^N  \Omega_{l} \mbox{\boldmath $\xi$}_l^{k} \right] k_1 t \right) &  \ldots & \left( \mbox{\boldmath $\eta$}_N^{k} +
\left[ \displaystyle\sum_{l=1}^N  \Omega_{l} \mbox{\boldmath $\xi$}_l^{k} \right] k_N t \right)
\end{array} \right],
\label{eq:matix _D2a}
\end{equation}
where we have considered $\prod_{i=1}^k \|\vec{w}_i(t)\| \propto
t^k$. Then, Eq.~(\ref{eq:norm}) can be used for the computation of
GALI$_k$.

In order to determine the leading order behavior of GALI$_k$ as $t$
grows, we look for the fastest increasing determinants of all $k
\times k$ minors of matrix $\textbf{W}^{k}$. For $2\leq k \leq 2N-1$,
these determinants include only one column of $\textbf{W}^{k}$
containing the term $\left[\sum_{l=1}^N \Omega_{l} \mbox{\boldmath
    $\xi$}_l^{k} \right]$ and grow proportional to $t$, since
determinants with more than one columns proportional to
$\left[\sum_{l=1}^N \Omega_{l} \mbox{\boldmath $\xi$}_l^{k} \right]$
are identically zero.  Thus, we conclude that $\mbox{GALI}_k (t)
\propto t^{-(k-1)}$ for $2 \leq k \leq 2N-1$. For $k=2N$,
$\textbf{W}^{k}$ is a square $2N \times 2N$ matrix which has a
constant determinant, since time appears only through multiplications
with the $N$ first columns of $\textbf{W}^{k}$, and so $\mbox{GALI}_k
(t) \propto t^{-2N}$. Summarizing, the time evolution of GALI$_k$ for
stable periodic orbits of $N$dof Hamiltonian systems is given by
\begin{equation}
\mbox{GALI}_k \propto \left\{ \begin{array}{ll} \frac{1}{t^{k-1}} & \mbox{if
$2 \leq k \leq 2N-1$} \\ \frac{1}{t^{2N}} & \mbox{if $ k = 2N$} \\
\end{array}\right..
\label{eq:GALI_PO_Ham}
\end{equation}
It is worth mentioning that Eq.~(\ref{eq:GALI_PO_Ham}) can be
retrieved from Eq.~(\ref{eq:GALI_order_all}) by assuming motion on an
$s=1$-dimensional torus, i.e.~on an 1-dimensional curve, which is the
stable periodic orbit. Note that for $s=1$, only the last two branches
of Eq.~(\ref{eq:GALI_order_all}) are meaningful.

Stable periodic orbits of symplectic maps correspond to stable fixed
points of the map, which are located inside islands of stability. Any
deviation vector from the stable periodic orbit performs a rotation
around the fixed point. This, for example, can be easily seen in the
case of 2D maps where the islands in the vicinity of a stable fixed
point can be represented through linearization, by ellipses (see for
instance \cite[Sect.~3.3.b]{LichtenbergL_1992}\cite{LF_01}). Thus, any
set of $2\leq k \leq 2N$ initially linearly independent, unit
deviation vectors will rotate around the fixed point, keeping on the
average the angles between them constant. This means that the volume
of the $k$-parallelogram having as edges these vectors will remain
practically constant, exhibiting some fluctuations, since the rotation
angles are constant only on average. So, in the case of stable
periodic orbits of $2N$D maps we have
\begin{equation}
\mbox{GALI}_k \propto \mbox{constant}, \,\,\, 2 \leq k \leq 2N.
\label{eq:GALI_PO_maps}
\end{equation}

\subsection{Numerical results - Hamiltonian flows}
\label{Hamflows}
To verify the validity of the theoretical predictions of
Eqs.~(\ref{eq:GALI_chaos_upo}) and (\ref{eq:GALI_PO_Ham}) we now
compute the GALIs for some representative Hamiltonian systems of
different number of degrees of freedom.

\subsubsection{2dof H\'{e}non-Heiles system}
\label{sect:2dof}

\begin{figure}[!ht]
\centering
  \includegraphics[width=5.1cm]{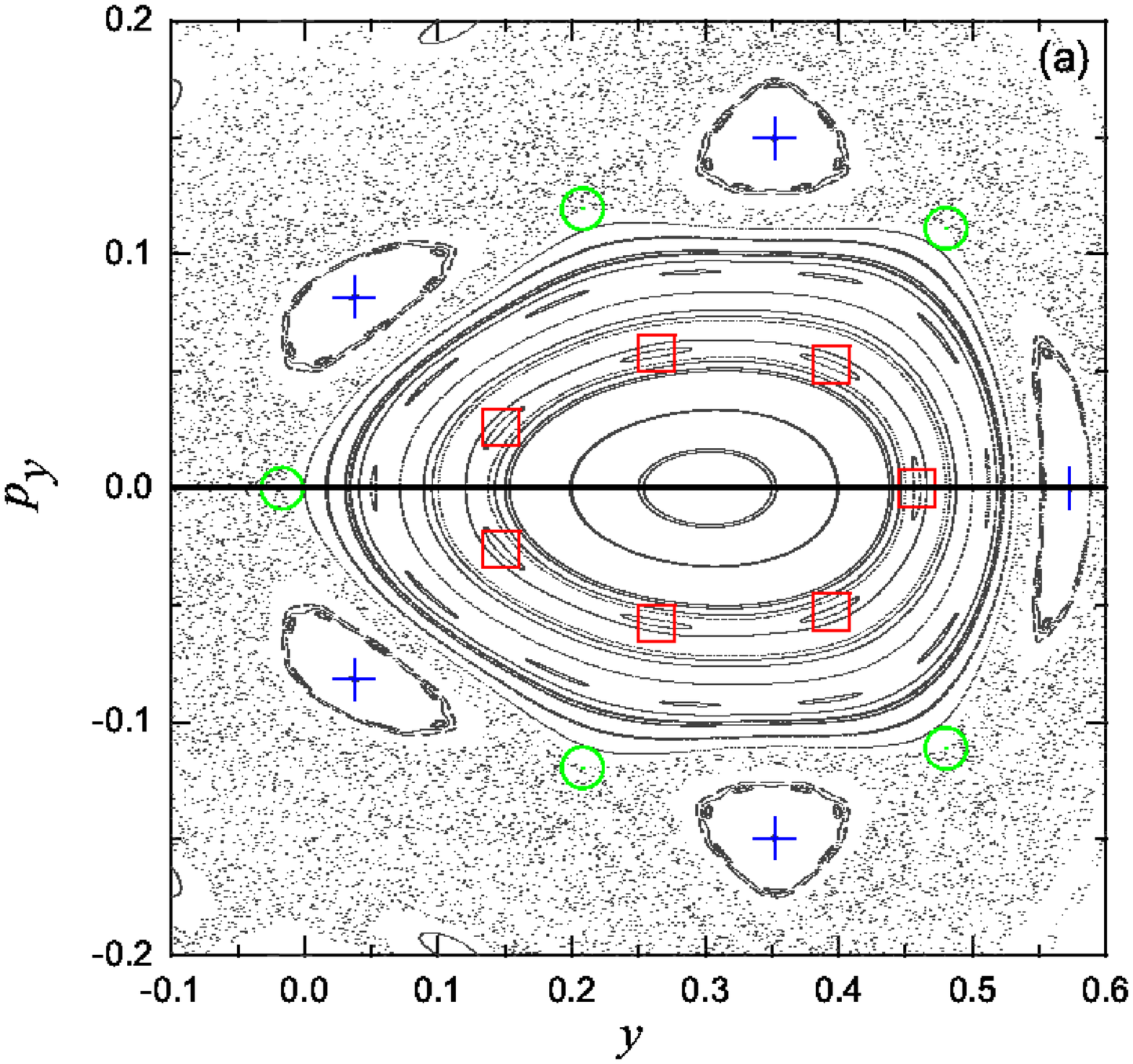}
  \includegraphics[width=5.cm]{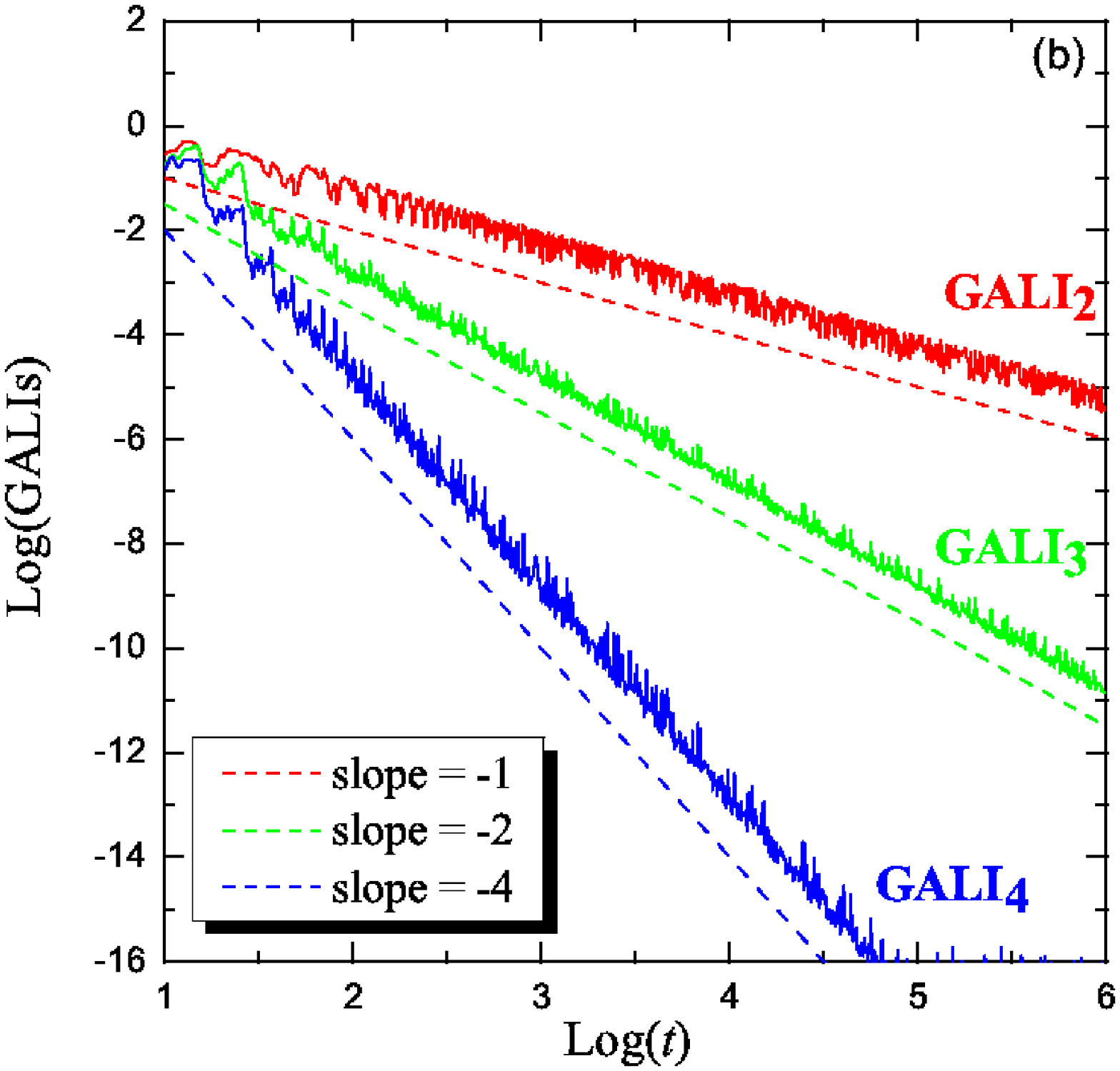}
  \includegraphics[width=5.cm]{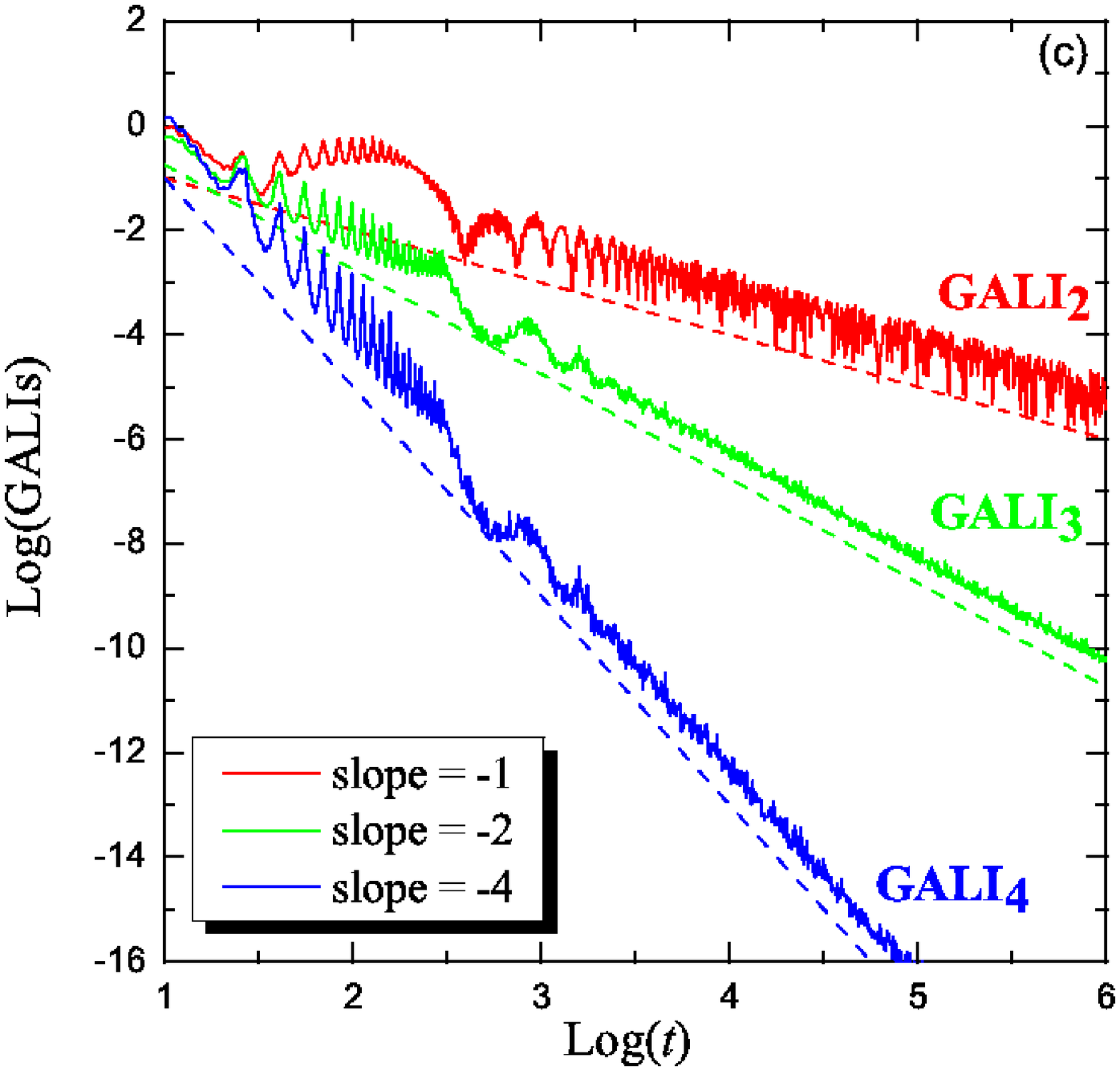}  \\
\vspace{1cm}
  \includegraphics[width=5.cm]{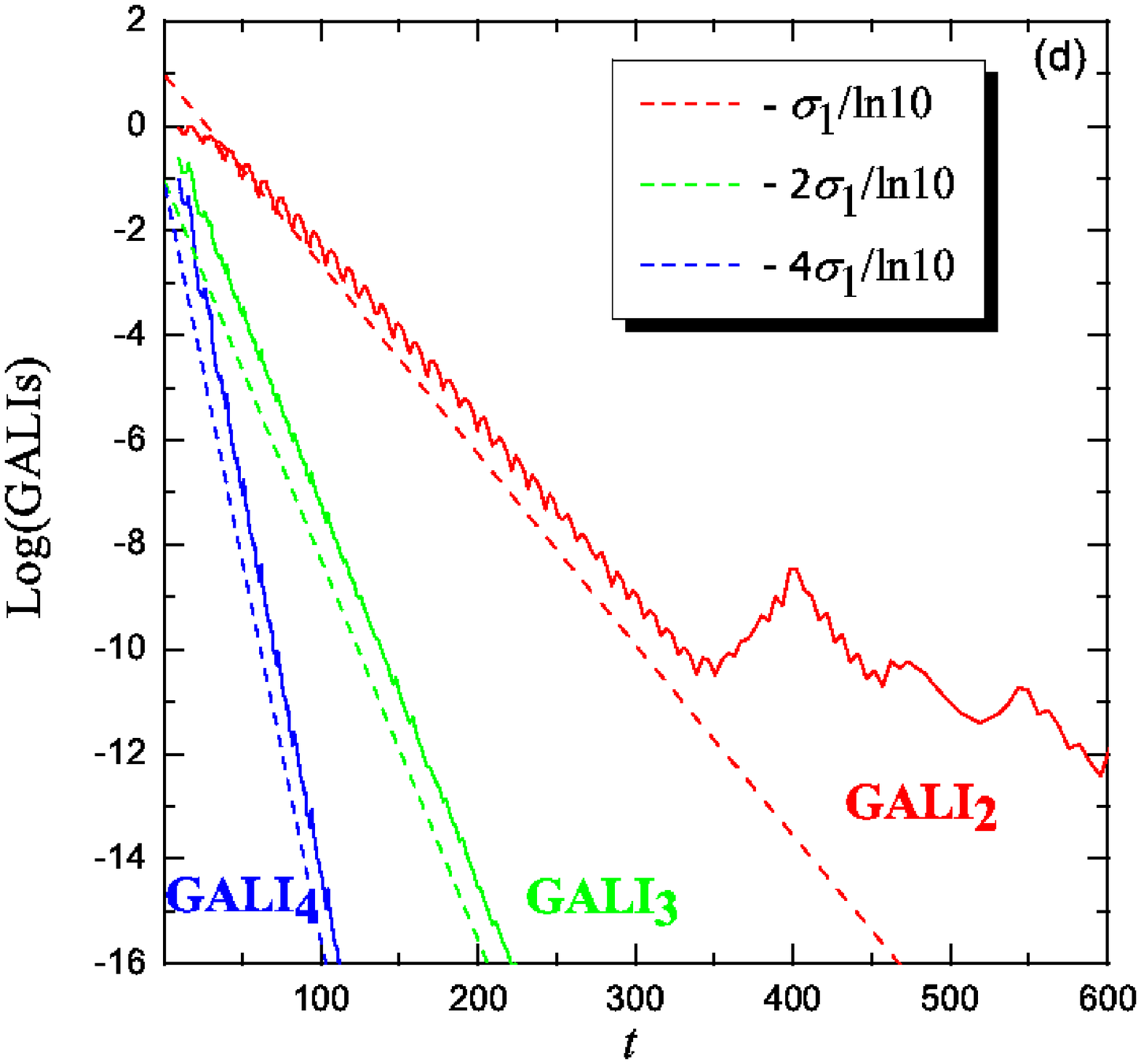}
  \includegraphics[width=5.cm]{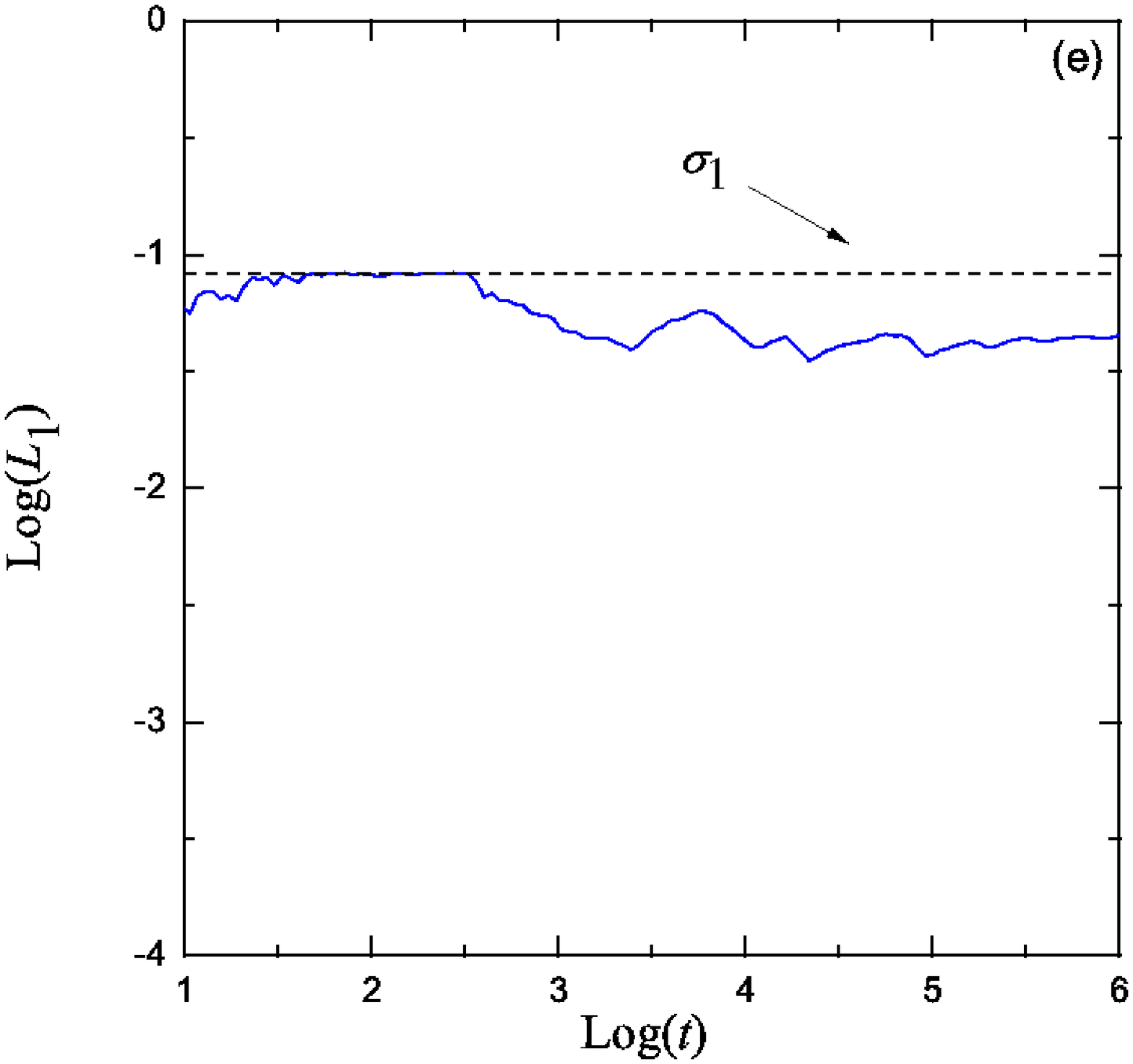}
  \caption{(a) The PSS of the 2dof H\'{e}non-Heiles system
    \eqref{2DHH} with $H_2=0.125$. The intersection points of stable
    periodic orbits of period 5 (blue crosses) and 7 (red squares), as
    well as an unstable orbit of period 5 (green circles) are also
    plotted. The line $p_y=0$, denoting a set of initial conditions
    discussed in Sect.~\ref{sec:near}, is also plotted. The time
    evolution of GALI$_2$ (red curves), GALI$_3$ (green curves) and
    GALI$_4$ (blue curves) for these three orbits is shown in panels
    (b), (c) and (d) respectively. Both axes of (b) and (c), and the
    vertical axis of (d) are logarithmic. (e) The time evolution of
    the quantity $L_1(t)$, which has as limit for $t\rightarrow
    \infty$ the mLCE $\sigma_1$ of the unstable periodic orbit
    (horizontal dotted line). Plotted lines correspond to functions
    proportional to $t^{-1}$, $t^{-2}$ and $t^{-4}$ in (b) and (c),
    and the exponential laws (\ref{eq:GALI_chaos_upo}) for $\sigma_1 =
    0.084$, $\sigma_2 = 0$ in (d).}
\label{2D_Ham_HH1}
\end{figure}

First we consider the well-known 2dof H\'{e}non-Heiles model \cite{HH}
\begin{equation}\label{2DHH}
    H_2=\frac{1}{2}(p_{x}^{2}+p_{y}^{2})+\frac{1}{2}(x^{2}+y^{2}) +
    x^{2}y-\frac{1}{3}y^{3}.
\end{equation}
In our study we keep the value of the Hamiltonian fixed at
$H_2=0.125$. Fig.~\ref{2D_Ham_HH1}(a) shows the PSS of the system
defined by $x=0$, $p_x \geq 0$. We consider two stable periodic orbits
(whose stability type is $S_1$ according to Eq.~(\ref{eq:sta_type})):
An orbit of period 5 (i.e.~an orbit intersecting the PSS at the 5
points denoted by blue crosses in Fig.~\ref{2D_Ham_HH1}(a)) with
initial condition $(x,y,p_x,p_y) \approx
(0.0,0.35207,0.36427,0.14979)$, and an orbit of period 7 (red squares
in Fig.~\ref{2D_Ham_HH1}(a)) with initial condition
$(x,y,p_x,p_y)\approx (0.0,0.45882,0.32229,0.0)$. The time evolution
of GALI$_k$, $k=2,3,4$ for these two orbits, for a random choice of
initial orthonormal deviation vectors, is shown in
Figs.~\ref{2D_Ham_HH1}(b) and (c) respectively. For both orbits the
indices show a power law decay to zero, in accordance with the
theoretical prediction of Eq.~(\ref{eq:GALI_PO_Ham}) for $N=2$.

In order to check the validity of Eq.~(\ref{eq:GALI_chaos_upo}), we
consider an unstable periodic orbit (of $U_1$ type) of period 5 (green
circles in Fig.~\ref{2D_Ham_HH1}(a)) with initial condition
$(x,y,p_x,p_y)\approx
(0.0,0.2083772012,0.4453146996,0.1196065752)$. The theoretically
expected value of this orbit's mLCE $\sigma_1$ is estimated from
Eq.~(\ref{eq:LCE_PO}) to be $\sigma_1\approx0.084$, while $\sigma_2=0$
because the Hamiltonian function is an integral of motion.

In Fig.~\ref{2D_Ham_HH1}(d) the time evolution of the corresponding
GALI$_k$, $k=2,3,4$ is plotted. From these results we conclude that
the computed values of GALIs are well approximated by
Eq.~(\ref{eq:GALI_chaos_upo}) for $\sigma_1 = 0.084$ and $\sigma_2=0$,
at least up to $t\approx 350$. After that time we observe a change in
the exponential decay of GALI$_2$. This happens because the
numerically computed orbit deviates from the unstable periodic orbit,
due to computational inaccuracies, and enters the surrounding chaotic
domain, which is characterized by different LCEs. This behavior is
also evident from the evolution of the finite time mLCE $L_1(t)$
(Fig.~\ref{2D_Ham_HH1}(e)) having as limit for $t\rightarrow \infty$
the mLCE $\sigma_1$ of the computed orbit (for more details on the
computation of the mLCE the reader is referred to
\cite[Sect.~5]{Sko:LE}). For an initial time interval, $L_1(t)$ well
approximates the mLCE of the unstable periodic orbit, but later on,
due to the divergence of the computed orbit from the periodic
trajectory, $L_1(t)$ tends to a different value, which is the mLCE of
the chaotic domain around the unstable periodic orbit.

\subsubsection{A 3dof Hamiltonian system}
\label{sect:3dof}

Let us now investigate the behavior of the GALIs for a 3dof
Hamiltonian system, where different types of unstable periodic orbits
can appear. In particular, we consider a system of three harmonic
oscillators with nonlinear coupling, described by the Hamiltonian
\begin{equation}
  H_3=\frac{1}{2}(p_{x}^{2}+p_{y}^{2}+p_{z}^{2})+\frac{1}{2}(Ax^{2}+By^{2}+Cy^{2})
  - \varepsilon xz^{2}- \eta yz^{2}.
\label{3DHam}
\end{equation}
The harmonic frequencies of the oscillators are determined by
parameters $A$, $B$, $C$, and the strengths of the nonlinear couplings
by $\varepsilon$ and $\eta$. This system was introduced as a crude
description of the inner parts of distorted 3-dimensional elliptic
galaxies. Detailed studies of its basic families of periodic orbits
were performed in
\cite{ConBar:1985,ConMag:1985,Con:1986a,Con:1986b}. Following these
works, we fix $A=0.9$, $B=0.4$, $C=0.225$ and $H_3=0.00765$ and vary
$\varepsilon$ and $\eta$ in order to study periodic orbits of
different stability types.

\begin{figure}[!ht]
\centering
  \includegraphics[width=4.9cm]{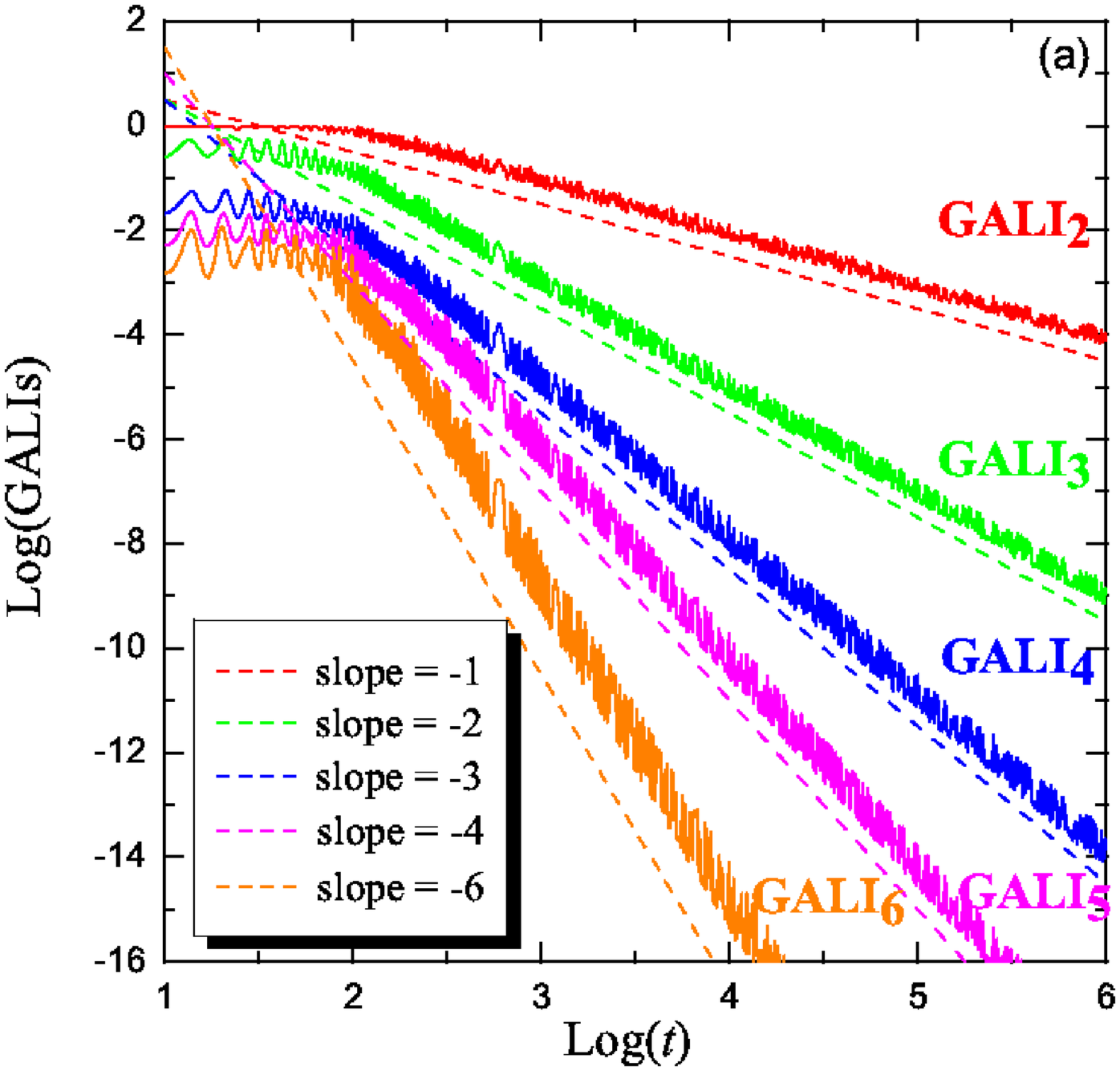}
  \includegraphics[width=5.cm]{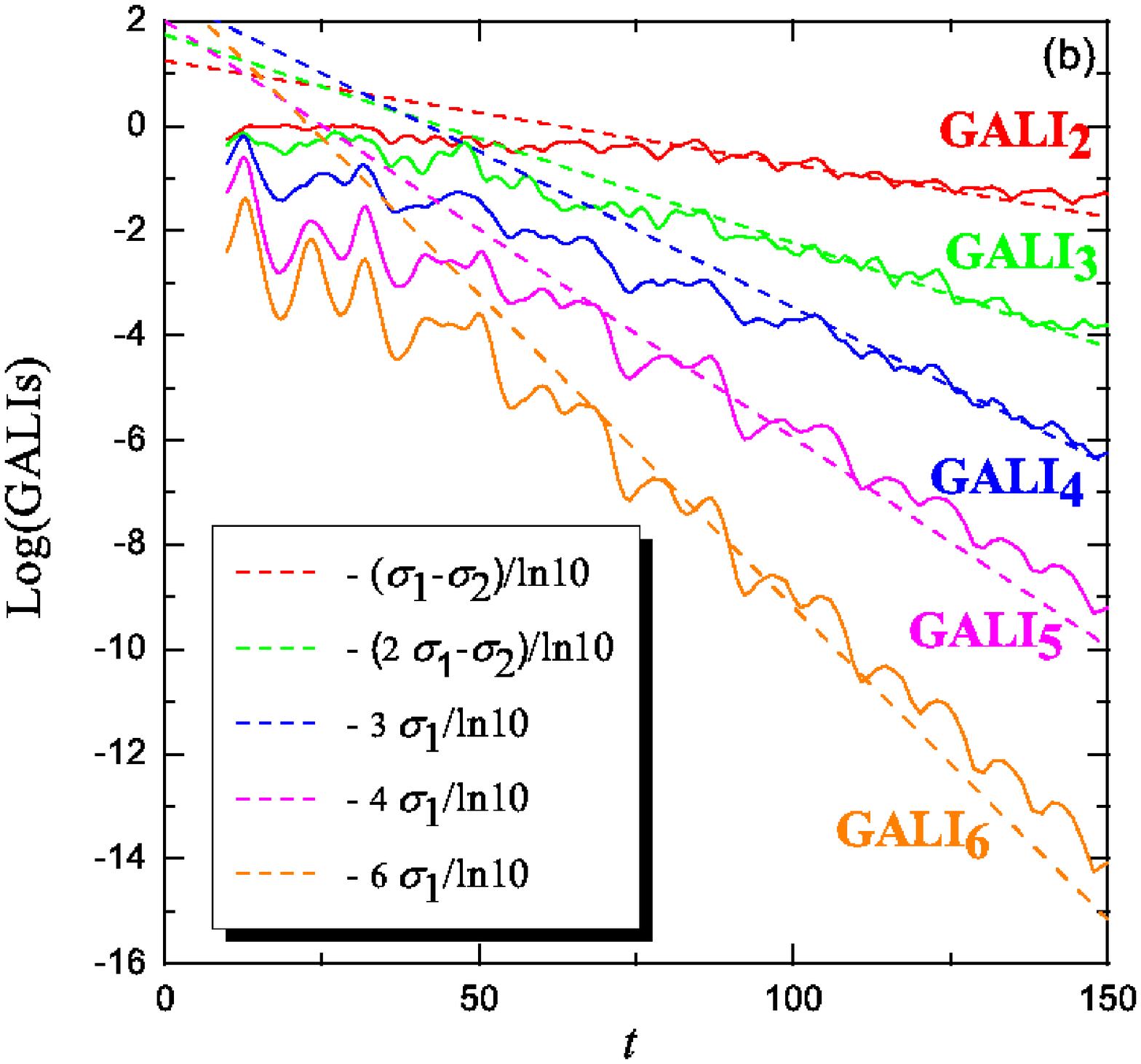}
  \includegraphics[width=5.cm]{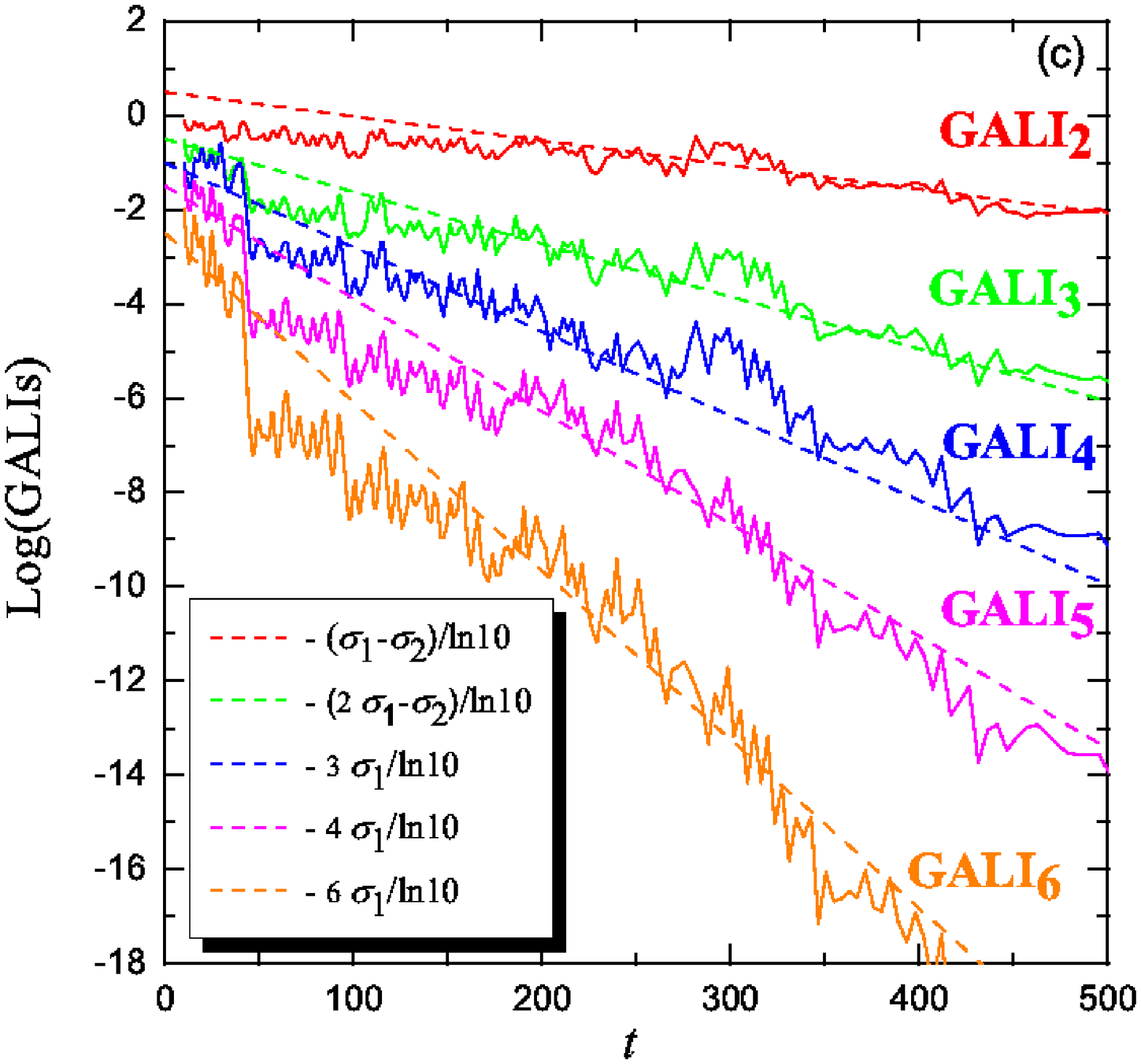}\\
\vspace{1cm}
  \includegraphics[width=5.cm]{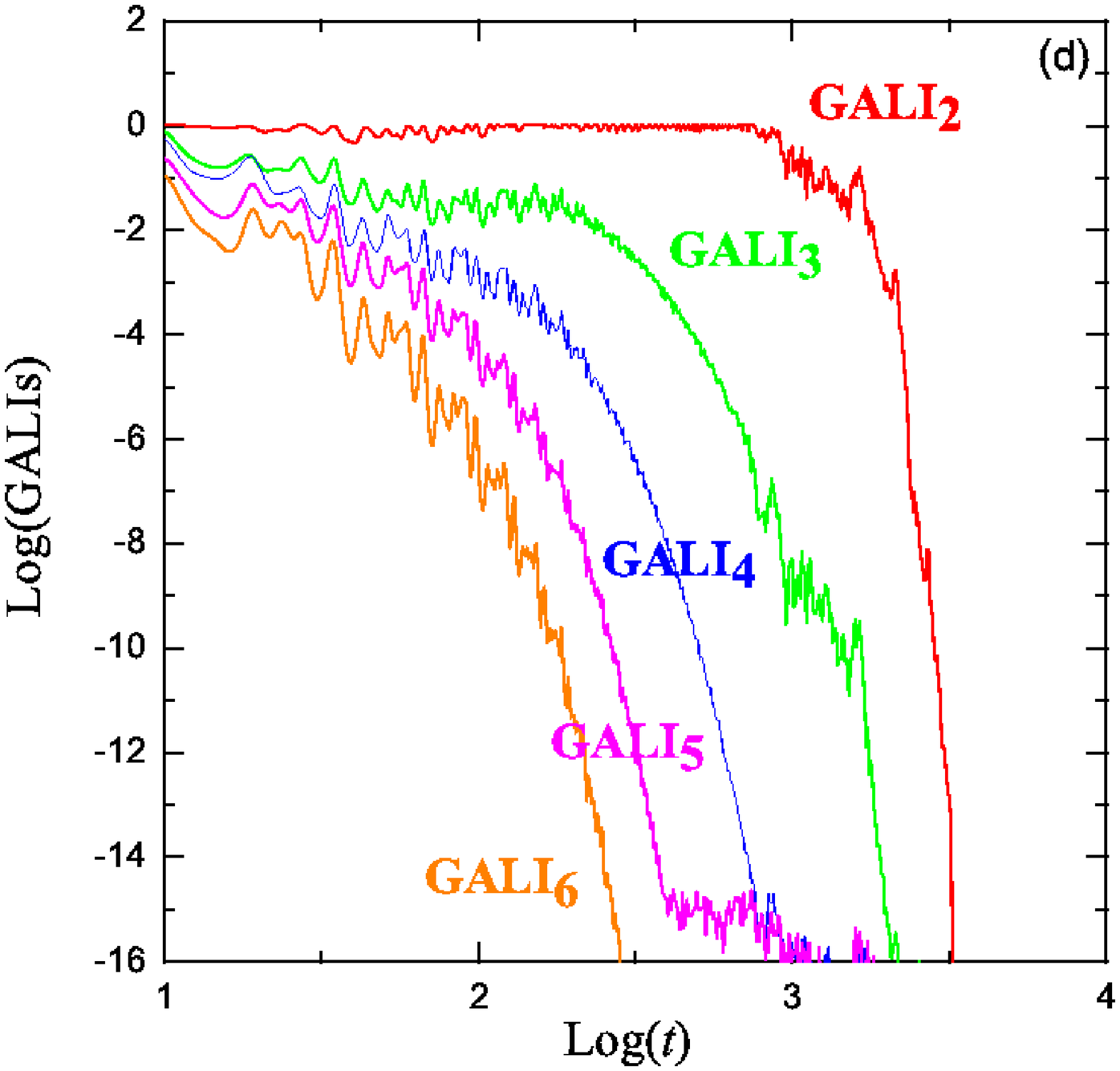}
  \includegraphics[width=5.cm]{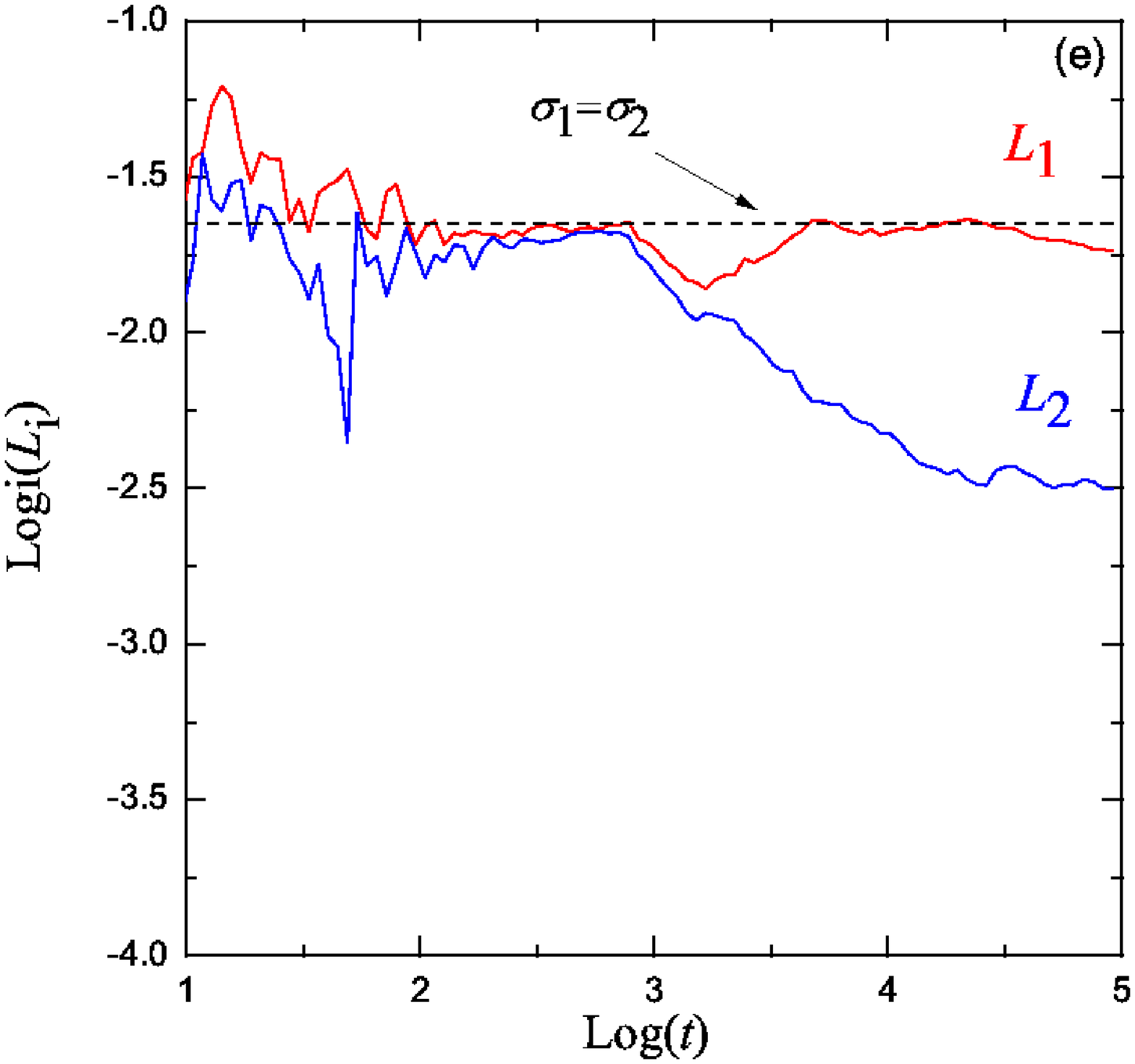}
  \caption{The time evolution of GALI$_k$, $2\leq k \leq 6$ for an (a)
    $S_2$ stable, (b) $S_1 U_1$ unstable, (c) $U_2$ unstable and (d)
    $\Delta_1$ unstable periodic orbit of the 3dof Hamiltonian system
    (\ref{3DHam}). Both axes of (a) and (d), and the vertical axis of
    (b) and (c) are logarithmic. Plotted lines correspond to
    appropriate power laws (\ref{eq:GALI_PO_Ham}) in (a), and
    exponential laws (\ref{eq:GALI_chaos_upo}) in (b) and (c). (e) The
    time evolution of quantities $L_1(t)$, $L_2(t)$ having
    respectively, as limit the two largest LCEs $\sigma_1$, $\sigma_2$
    of the $\Delta_1$ unstable periodic orbit. The theoretically
    estimated value $\sigma_1=\sigma_2 = 0.023$ is denoted by a
    horizontal line.}
\label{SPOUPO}
\end{figure}

In Fig.~\ref{SPOUPO}(a) we plot the time evolution of GALIs for a
stable ($S_2$) periodic orbit with initial condition
$(x,y,z,p_x,p_y,p_z)\approx (-0.06686,0.01230,0,0,0,0.10590)$ for
$\varepsilon=0.2$ and $\eta=0.1$. The 3dof system has a 6-dimensional
phase space and so, 5 different GALI$_k,$ with $2\leq k \leq 6$, are
defined. All GALIs decay to zero following the power law predictions
given by Eq.~(\ref{eq:GALI_PO_Ham}) for $N=3$.

Let us now study representative cases of all the different types of
unstable periodic orbits that can appear in a general 3dof system. In
particular, we consider an $S_1 U_1$ periodic orbit with initial
condition $(x,y,z,p_x,p_y,p_z)\approx
(-0.0238841214,0744533850,0,0,0,0.1121127613)$ for
$\varepsilon=0.848$, $\eta=0.1$ (Fig.~\ref{SPOUPO}(b)), an $U_2$
periodic orbit with initial condition
$(x,y,z,p_x,p_y,p_z)\approx(-0.0392937629,0.0648373644,0,-0.0564496390,0.0021636015,0.0950663122)$
for $\varepsilon=0.35$, $\eta=0.51$ (Fig.~\ref{SPOUPO}(c)), and a
$\Delta_1$ periodic orbit with initial condition
$(x,y,z,p_x,p_y,p_z)\approx(-0.0456720106,0.0658047594,0,0,0,0.1081228661)$
for $\varepsilon=0.6$ and $\eta=0.3$ (Figs.~\ref{SPOUPO}(d) and (e)).

Using Eq.~(\ref{eq:LCE_PO}) we estimated the LCEs to be $\sigma_1
\approx 0.046$, $\sigma_2=0$ and $\sigma_1\approx 0.014$,
$\sigma_2\approx 0.0019$ for the $S_1 U_1$ and the $U_2$ unstable
periodic orbits respectively. Using these values as good
approximations of the actual LCEs, we see in Figs.~\ref{SPOUPO}(b) and
(c) that the evolution of GALIs is well reproduced by
Eq.~(\ref{eq:GALI_chaos_upo}).

An eigenvalue of the monodromy matrix of the $\Delta_1$ unstable
periodic orbit is numerically found to be $\lambda_1 \approx
1.410+0.164i$, while the remaining three of them (apart from the two
unit ones) are $1/\lambda_1$, $\lambda_1^*$ and $1/\lambda_1^*$. Then,
from Eq.~(\ref{eq:LCE_PO}) we estimated the three largest LCEs of the
periodic orbit to be $\sigma_1=\sigma_2\approx0.023$,
$\sigma_3=0$. The evolution of the GALIs for this orbit is shown in
Fig.~\ref{SPOUPO}(d). Although the periodic orbit is unstable,
GALI$_2$ does not decay to zero but remains constant until $t\approx
10^3$. This happens because, according to
Eq.~(\ref{eq:GALI_chaos_upo}) GALI$_2 \propto
e^{-(\sigma_1-\sigma_2)t}$, but in this case $\sigma_1=\sigma_2$.
However, due to unavoidable inaccuracies in the numerical integration,
the computed orbit eventually diverges from the unstable periodic one
and enters a chaotic domain characterized by different LCEs with
$\sigma_1 \neq \sigma_2$. This divergence is also evident from the
evolution of quantities $L_1(t)$, $L_2(t)$ in Fig.~\ref{SPOUPO}(e),
whose limits at $t\rightarrow \infty$ are $\sigma_1$ and $\sigma_2$
respectively (see \cite{Sko:LE} for more details on the computation of
$\sigma_1$ and $\sigma_2$). In particular, we get $L_1(t) \approx
L_2(t)$ for $t \lesssim 10^3$, while later on the two quantities
attain different values. Consequently, for $t\gtrsim 10^3$ GALI$_2$
starts to decay exponentially to zero. On the other hand, all other
GALIs in Fig.~\ref{SPOUPO}(d) show an exponential decay, even when
GALI$_2$ remains constant, since the corresponding exponents in
Eq.~(\ref{eq:GALI_chaos_upo}) do not vanish.

\subsubsection{A multi-dimensional Hamiltonian system}
\label{sect:5dof}

Finally, we turn to a multi-dimensional Hamiltonian system
representing a 1-dimensional chain of 5 identical particles with
nearest neighbor interactions given by the FPU-$\beta$ Hamiltonian
\cite{Fermi_1955}
\begin{equation}
  H_5=\frac{1}{2}\sum_{j=1}^{5}
  p_{j}^{2}+\sum_{j=0}^{5}\biggl(\frac{1}{2}(x_{j+1}-x_{j})^{2}+\frac{1}{4}\beta(x_{j+1}-x_{j})^{4}\biggr),
\label{FPU_Hamiltonian_2}
\end{equation}
where $x_{j}$ is the displacement of the $j$th particle from its
equilibrium position and $p_{j}$ is the corresponding conjugate
momentum. In our study, we set $\beta=1.04$ and impose fixed boundary
conditions to the system, so that we always have $x_0=x_6=0$.

\begin{figure}[!ht]
\centering
  \includegraphics[width=5.cm]{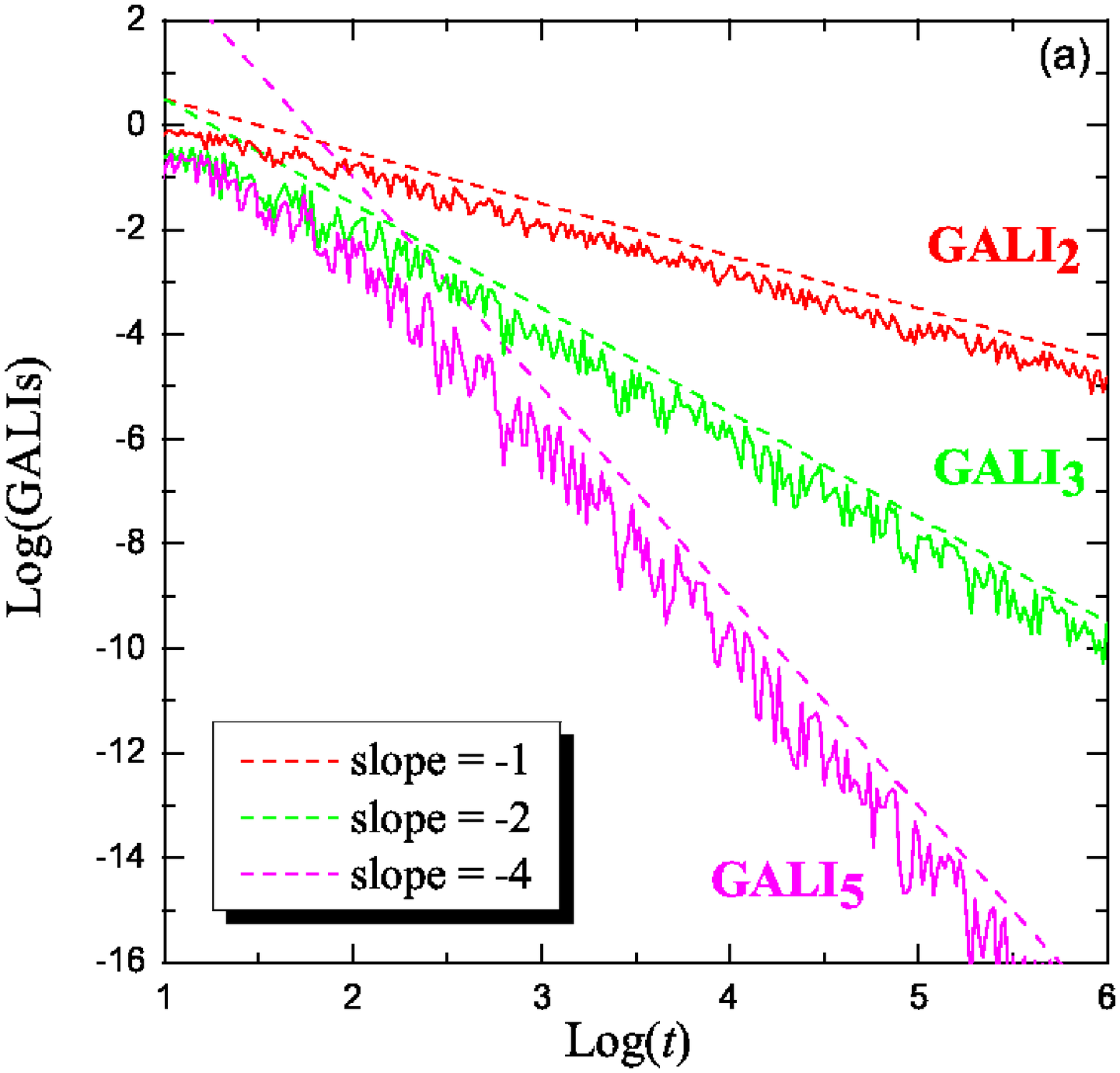}
  \includegraphics[width=5.cm]{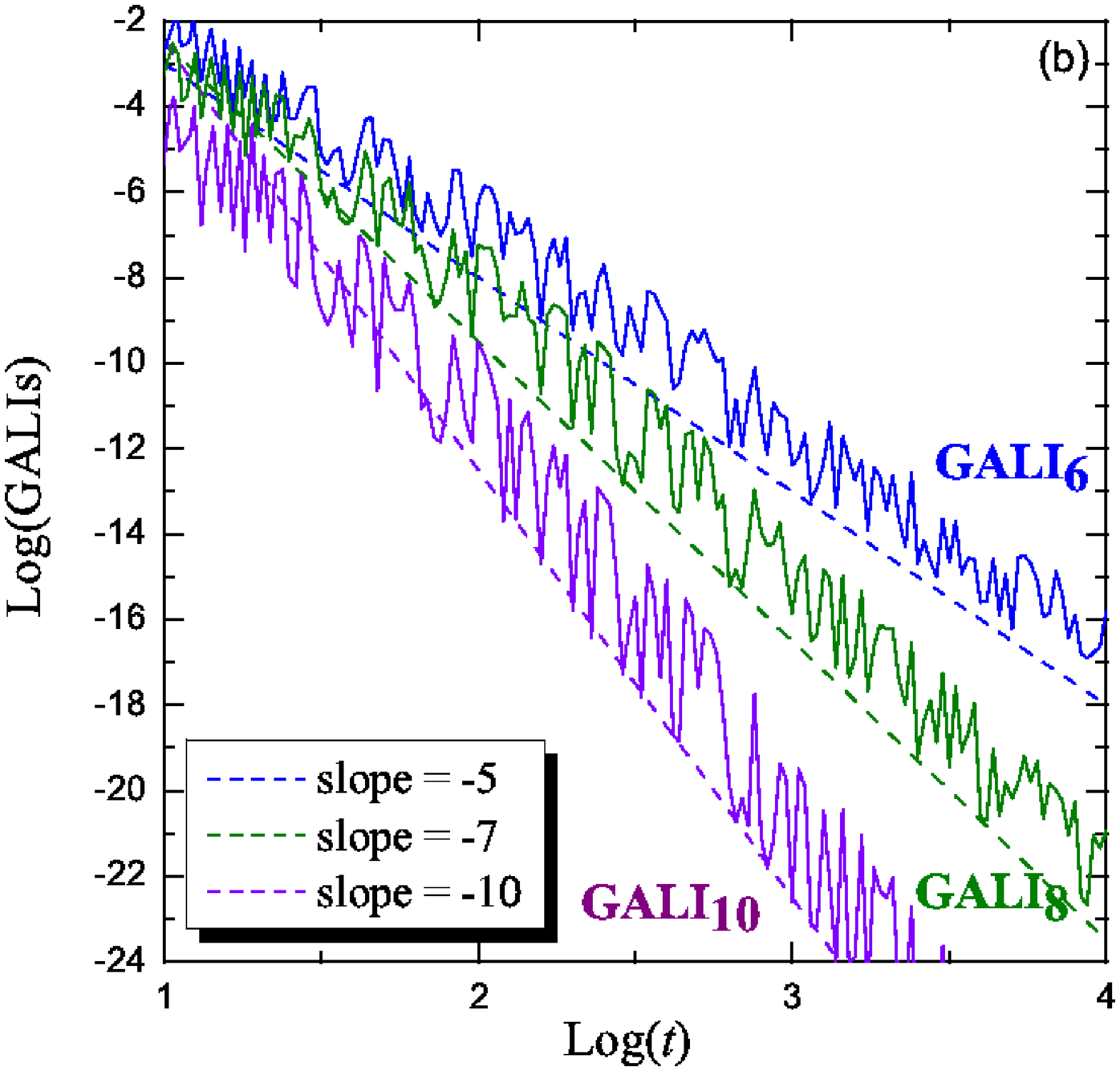}
  \includegraphics[width=5.cm]{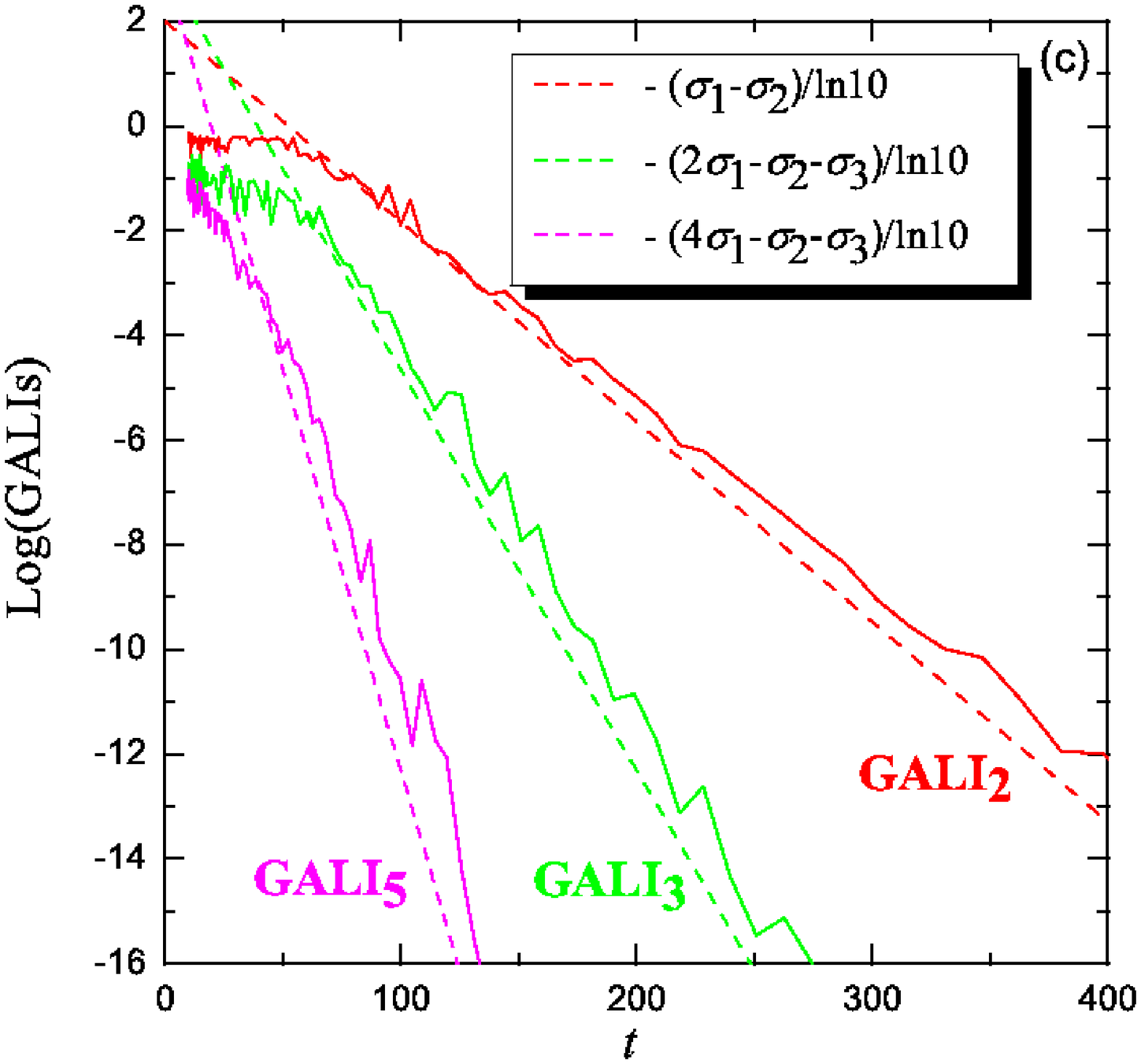}
  \caption{The time evolution of (a) GALI$_2$, GALI$_3$, GALI$_5$ and
    (b) GALI$_6$, GALI$_8$, GALI$_{10}$ for a stable periodic orbit of
    the 5dof Hamiltonian system (\ref{FPU_Hamiltonian_2}). (c) The
    time evolution of GALI$_2$, GALI$_3$, GALI$_5$ for an $S_4U_1$
    unstable periodic orbit of the same model. Plotted lines
    correspond to appropriate power laws (\ref{eq:GALI_PO_Ham}) in (a)
    and (b), and exponential laws (\ref{eq:GALI_chaos_upo}) in (c).}
\label{FigFPU:2}
\end{figure}

Let us consider two particular members of a family of periodic orbits
studied in \cite{Ooyama,Antonop_IJBC} which have initial conditions of
the form $x_1(0)=-x_3(0)=x_5(0)=\hat{x}(0)$, $x_2(0)=x_4(0)=0$,
$p_j(0)=0$, $1 \leq j \leq 5$. We compute the GALIs of an $S_5$ stable
periodic orbit (Figs.~\ref{FigFPU:2}(a) and (b)) with initial
condition $\hat{x}(0)\approx 1.035$ for $H_5=5$ and an $S_4U_1$
unstable periodic orbit (Fig.~\ref{FigFPU:2}(c)) with initial
condition $\hat{x}(0) \approx 1.168$ for $H_5=7$. From
Fig.~\ref{FigFPU:2} we see again that the behavior of the GALIs is
well reproduced by Eq.~(\ref{eq:GALI_PO_Ham}) for $N=5$ in the case of
the stable orbit, and by Eq.~(\ref{eq:GALI_chaos_upo}) for $\sigma_1
= 0.088$, $\sigma_i = 0$, $2\leq i \leq 5$, which are the values
obtained by Eq.~(\ref{eq:LCE_PO}) for the unstable orbit.

\subsection{Numerical results - Symplectic maps}
\label{SyMaps}

According to the theoretical arguments of Sect.~\ref{sec:theory}, the
GALIs of unstable periodic orbits of maps should exhibit the same
behavior as in the case of Hamiltonian flows, i.e.~they should tend
exponentially to zero following Eq.~(\ref{eq:GALI_chaos_upo}). On the
other hand, we have argued that the GALIs of stable periodic orbits
should remain constant, according to Eq.~(\ref{eq:GALI_PO_maps}),
having a different behavior with respect to Hamiltonian systems. To
verify these predictions, we now proceed to study some periodic orbits
in a 2D and a 4D symplectic map.

\subsubsection{2D H\'{e}non map}
\label{sect:2DHM}

\begin{figure}[!ht]
\centering
  \includegraphics[width=5.cm]{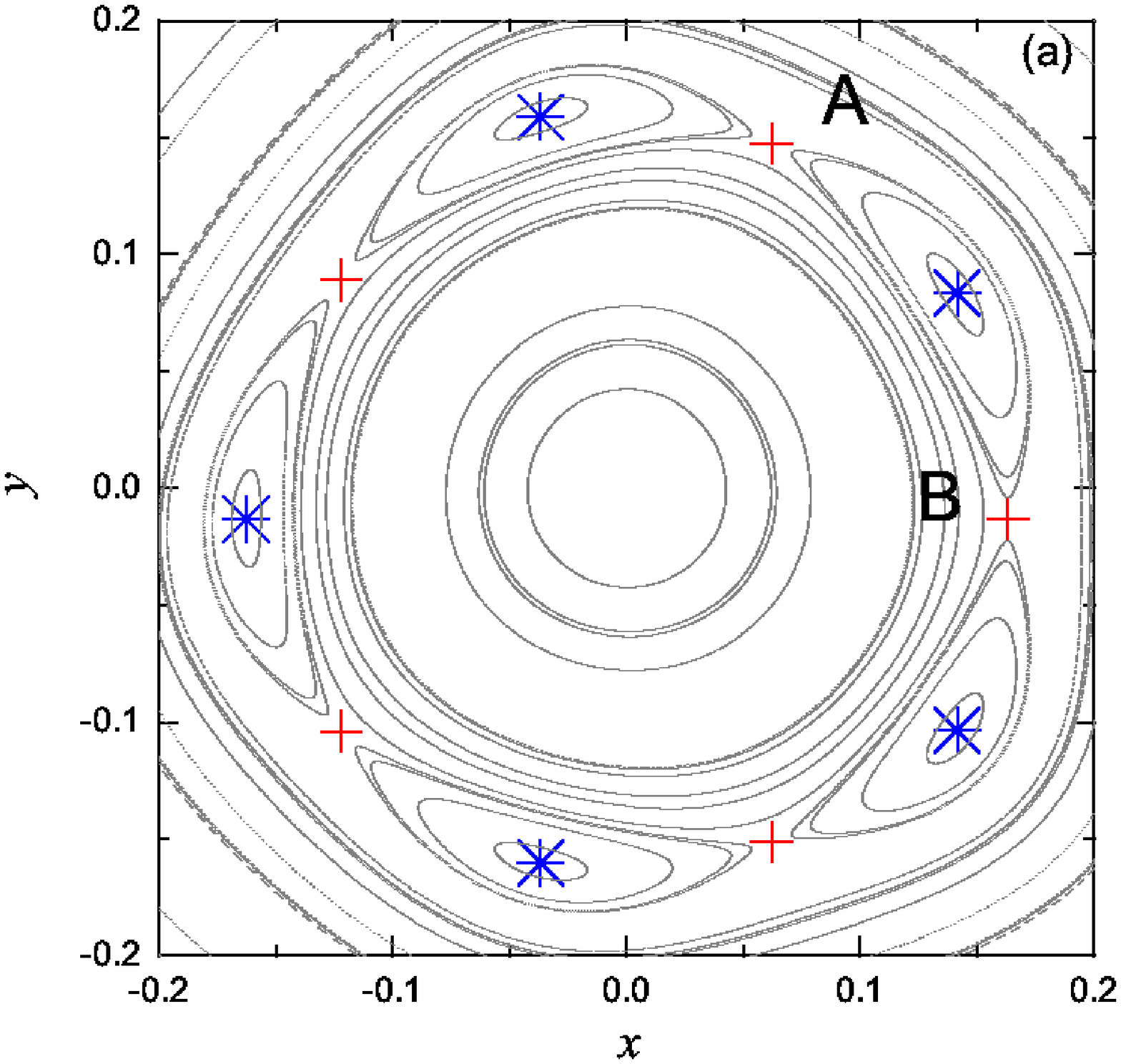}
  \includegraphics[width=5.cm]{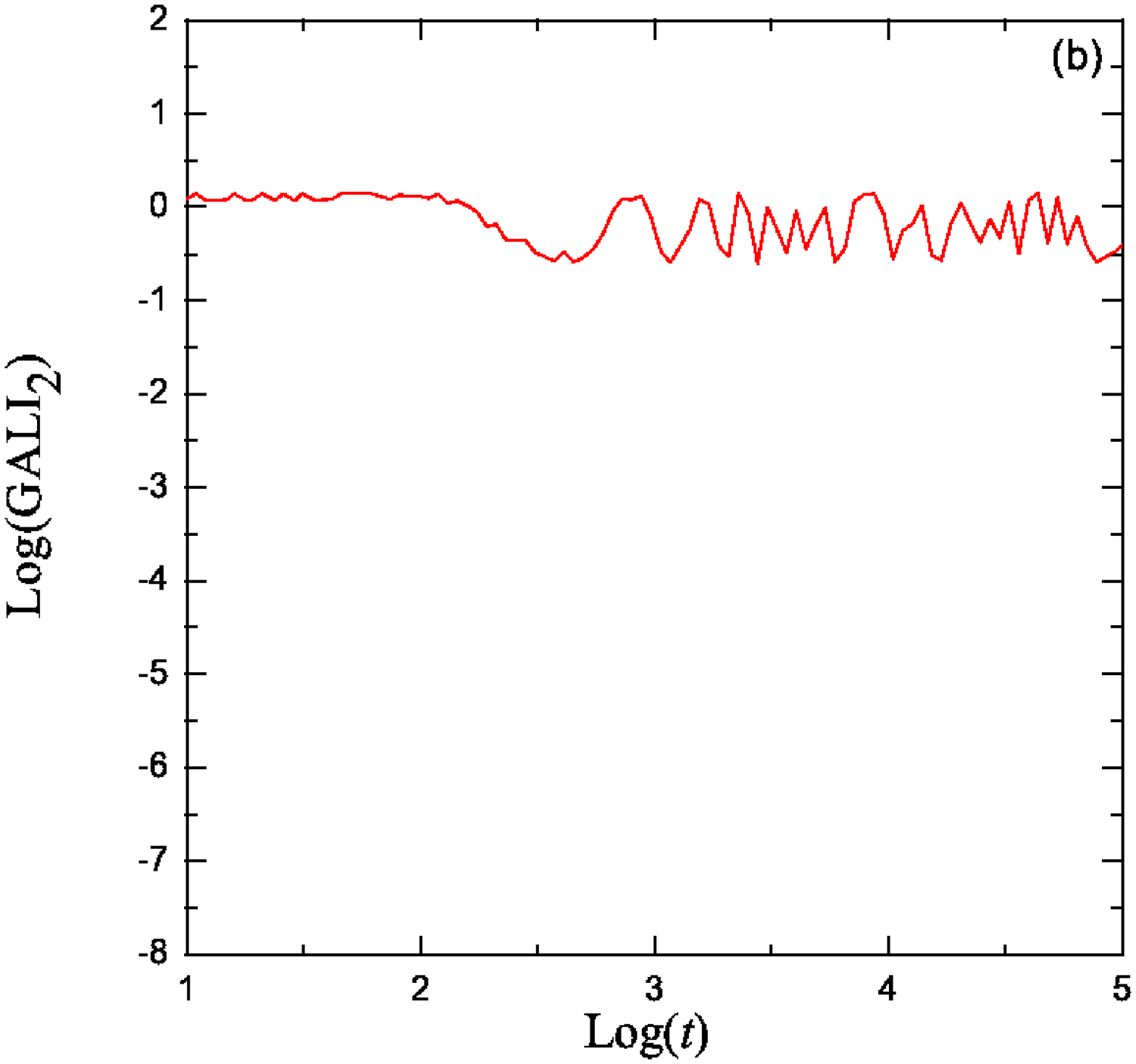}
  \caption{(a) The phase space of the 2D H\'{e}non map
    (\ref{2DHenMap}) for $\omega=0.201$. The points of two periodic
    orbits of period 5, a stable (blue stars) and an unstable one (red
    crosses), are also plotted. The time evolution of GALI$_2$ of the
    stable orbit is plotted in (b). Two particular points of the
    unstable periodic orbit discussed in Sect.~\ref{sec:near} are
    marked by letters A and B.}
\label{2DHM:1}
\end{figure}

First we consider the 2D H\'{e}non map \cite{Henon1969}
\begin{equation}
\begin{array}{lll}
  x^{'} &=&  x \cos(2\pi \omega) + (y+x^2) \sin(2\pi \omega)\\
  y^{'} &=& -x \sin(2\pi \omega) + (y+x^2) \cos(2\pi \omega),
\end{array}
\label{2DHenMap}
\end{equation}
where $\omega$ is a real positive constant. The phase space of this
map for $\omega=0.201$ is plotted in Fig.~\ref{2DHM:1}(a). We consider
two periodic orbits of period 5 (i.e.~after 5 iterations of the map
the orbit returns to its initial point): an $S_1$ stable orbit (blue
stars in Fig.~\ref{2DHM:1}(a)) with initial condition
$(x,y)\approx(0.14175,-0.10366)$, and an $U_1$ unstable one (red
crosses in Fig.~\ref{2DHM:1}(a)) with initial condition
$(x,y)\approx(0.0622148475,0.1477550294)$. Fig.~\ref{2DHM:1}(b) shows
that the GALI$_2$ of the stable periodic orbit oscillates around a
constant positive value, in accordance to
Eq.~(\ref{eq:GALI_PO_maps}). We have also verified that the GALI$_2$
of the unstable periodic orbit decays exponentially to zero following
Eq.~(\ref{eq:GALI_chaos_upo}) with $\sigma_1 = 0.0039$.

\subsubsection{4D standard map}
\label{sect:4DSM}

Let us now consider the 4D symplectic map \cite{Kan:1}
\begin{equation}\label{4Dmap}
 \begin{array}{lll}
  x_{1}' &=& x_{1} + x_{2}'\\
  x_{2}' &=& x_{2} + \frac{K_1}{2\pi}\sin(2\pi x_{1})-\frac{\beta}{2\pi}\sin[2\pi(x_{3}-x_{1})]\\
  x_{3}' &=& x_{3} + x_{4}'\\
  x_{4}' &=& x_{4} + \frac{K_2}{2\pi}\sin(2\pi x_{3})-\frac{\beta}{2\pi}\sin[2\pi(x_{1}-x_{3})]\\
 \end{array}
(\mbox{mod}\,\, 1),
\end{equation}
which consists of two coupled standard maps, with real parameters
$K_1$, $K_2$ and $\beta$.

\begin{figure}[!ht]
\centering
  \includegraphics[width=5.cm]{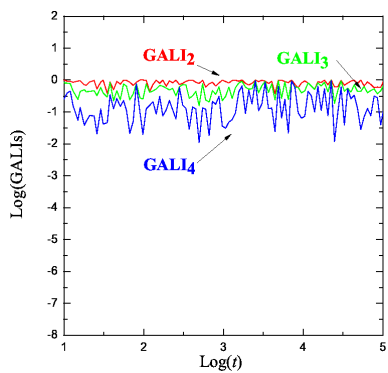}
  \caption{The time evolution of GALI$_2$ (red curve), GALI$_3$
    (green curve) and GALI$_4$ (blue curve) for a stable periodic
    orbit of period 7 of the 4D map (\ref{4Dmap}).}\label{4Dsm:1}
\end{figure}

In Fig.~\ref{4Dsm:1} we plot the evolution of GALIs for an $S_2$
stable periodic orbit of period 7 with initial condition
$(x_1,x_2,x_3,x_4)\approx(0.23666,0,0.23666,0)$ for $K_1=K_2=0.9$ and
$\beta=0.05$. Like in the case of the 2D map (\ref{2DHenMap}),
GALI$_2$, GALI$_3$ and GALI$_4$ remain constant, oscillating around
non-zero values, in accordance to Eq.~(\ref{eq:GALI_PO_maps}).

\section{Dynamics in the Neighborhood of Periodic Orbits}
\label{sec:near}

\begin{figure}[!ht]
\centering
  \includegraphics[width=5.cm]{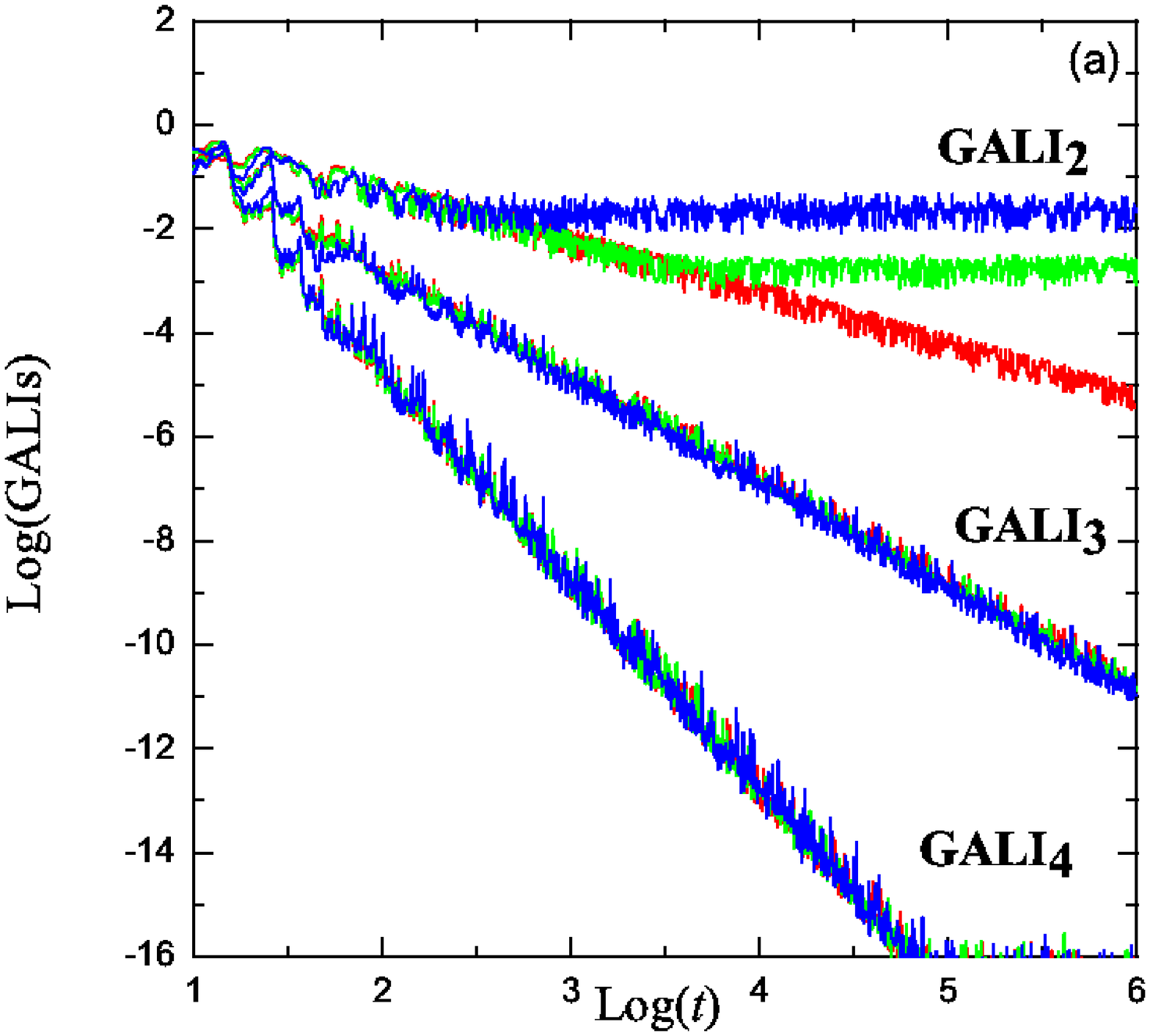}
  \includegraphics[width=5.08cm]{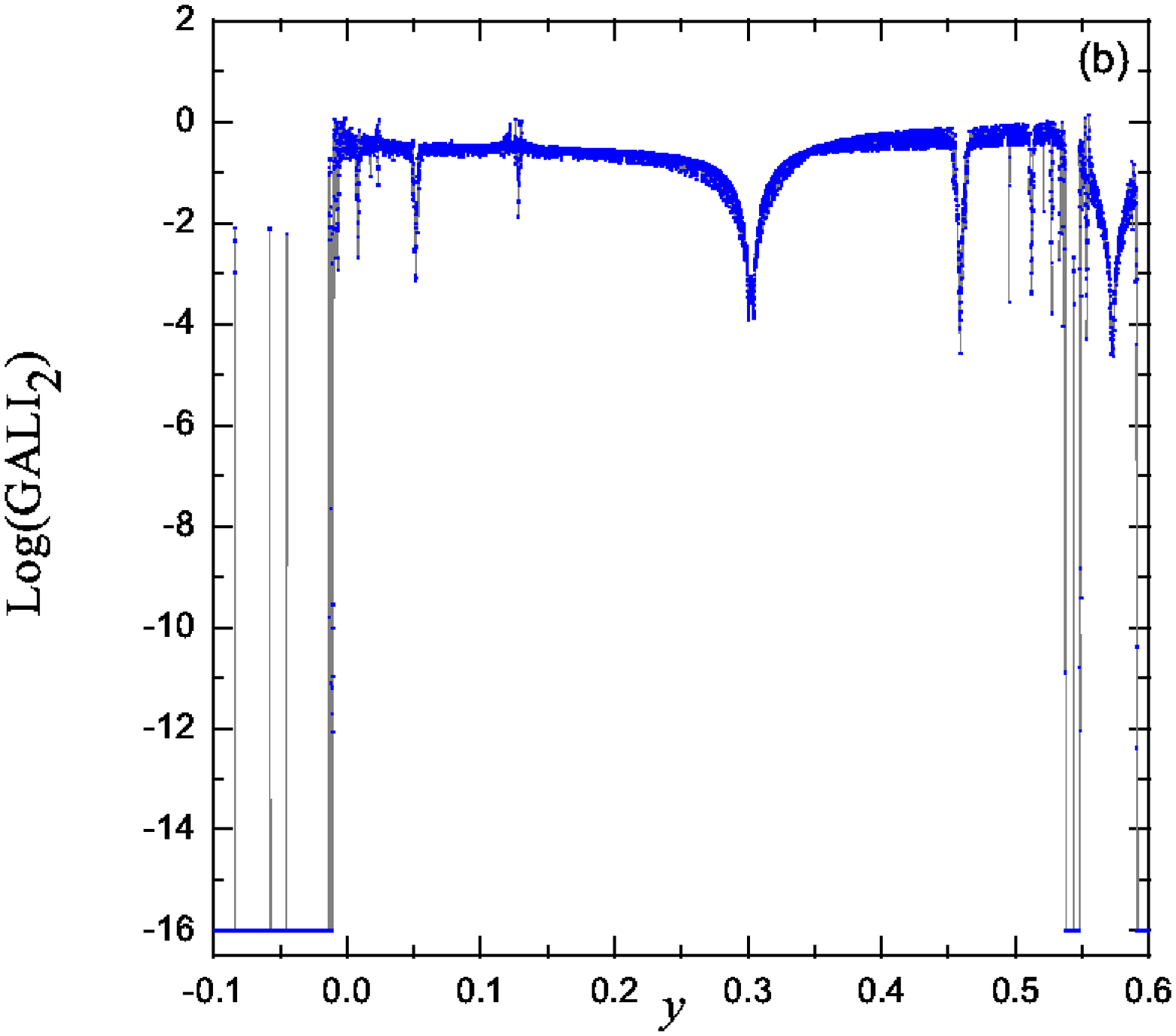}
  \includegraphics[width=5.3cm]{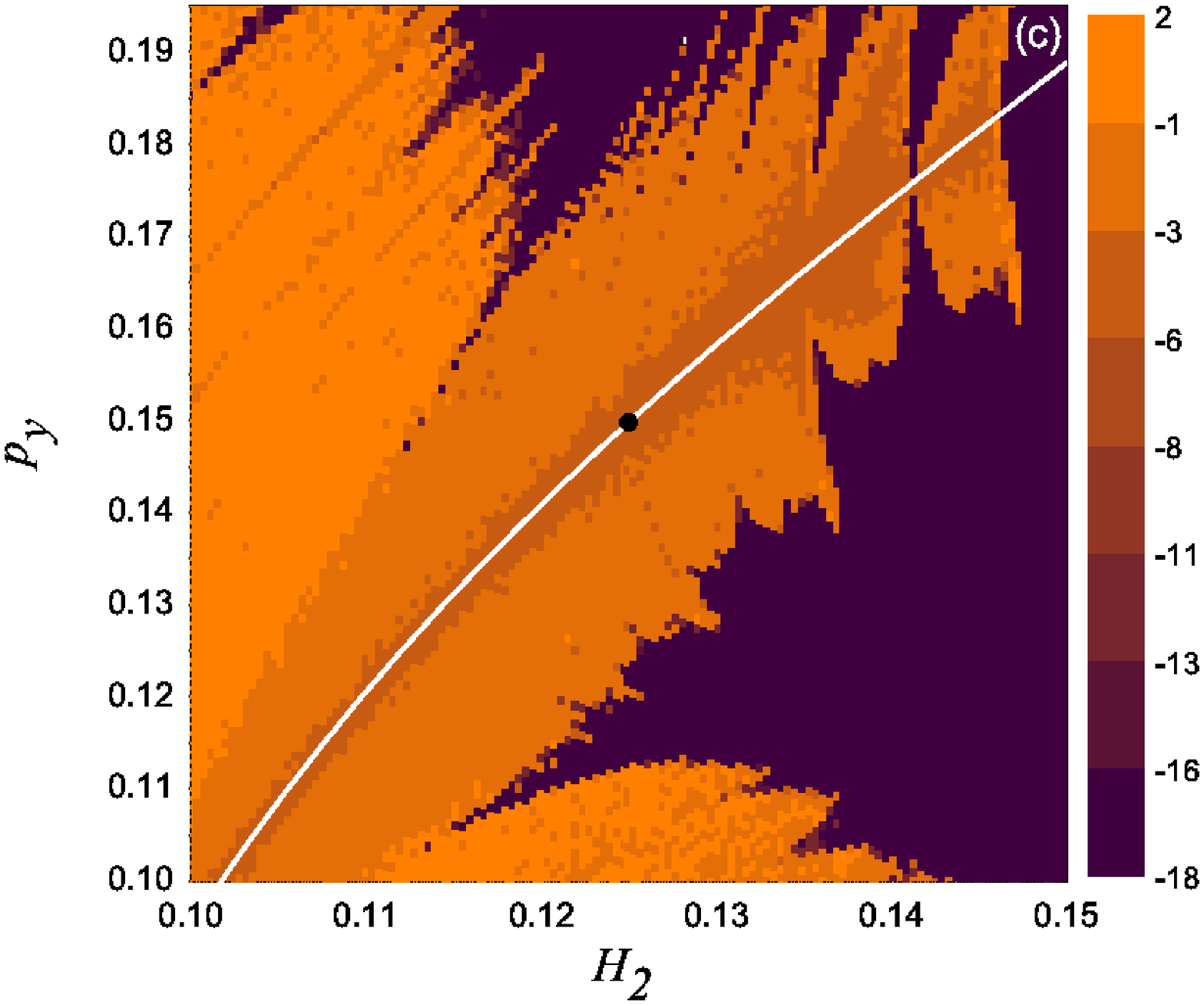}
  \caption{(a) The time evolution of GALI$_2$, GALI$_3$ and GALI$_4$
    for three orbits of the H\'{e}non-Heiles system (\ref{2DHH}): the
    stable periodic orbit of period 5 studied in Sect.~\ref{sect:2dof}
    (red curves) and two nearby orbits whose initial conditions result
    from a $\Delta y= 0.00793$ (green curves) and $\Delta y= 0.02793$
    (blue curves) perturbations of the periodic orbit. Note that
    curves of GALI$_3$ and GALI$_4$ overlap each other. (b) The
    GALI$_2$ values at $t=10^5$ for orbits with initial conditions on
    the $p_y=0$ line of the PSS of Fig.~\ref{2D_Ham_HH1}(a), as a
    function of the $y$ coordinate of the initial condition. (c)
    Regions of different GALI$_2$ values on the $(H_2,p_y)$ plane of
    the H\'{e}non-Heiles system (\ref{2DHH}). Each point corresponds
    to an orbit in the neighborhood of a family of periodic orbits
    (white curve) and is colored according to the
    $\log(\mbox{GALI}_2)$ value computed at $t=10^4$. The black filled
    circle denotes the stable periodic orbit of
    Fig.~\ref{2D_Ham_HH1}(b).}\label{nearSPO}
\end{figure}

We now turn our attention to the dynamics in the \textit{vicinity} of
periodic orbits, studying initially the neighborhood of stable
periodic orbits in Hamiltonian systems. As a first example we consider
the 2dof H\'{e}non-Heiles system (\ref{2DHH}), and in particular the
stable periodic orbit of period 5 studied in Sect.~\ref{sect:2dof}. In
Fig.~\ref{2D_Ham_HH1}(b) we have seen that GALI$_2 \propto t^{-1}$,
GALI$_3 \propto t^{-2}$ and GALI$_4 \propto t^{-4}$ in accordance to
Eq.~(\ref{eq:GALI_PO_Ham}). We expect that small perturbations of this
trajectory will lead to regular motion on 2-dimensional tori
surrounding the periodic orbit. For this kind of motion,
Eq.~(\ref{eq:GALI_order_all_N}) predicts GALI$_2 \propto$ constant,
GALI$_3 \propto t^{-2}$ and GALI$_4 \propto t^{-4}$.  Thus, only for
GALI$_2$ a different evolution between the periodic orbit and its
neighborhood is expected. This is actually true, as we see in
Fig.~\ref{nearSPO}(a) where the time evolution of GALI$_k$, $k=2,3,4$
is plotted for the stable periodic orbit (red curves) and two nearby
orbits whose initial conditions result from a $\Delta y= 0.00793$
(green curves) and $\Delta y= 0.02793$ (blue curves) perturbation. The
GALI$_2$ of neighboring orbits initially follows a GALI$_2 \propto
t^{-1}$ evolution, similar to the periodic orbit, but later on
stabilizes to a non-zero value as Eq.~(\ref{eq:GALI_order_all_N})
predicts. From Fig.~\ref{nearSPO}(a) we see that the closer the orbit
is to the periodic trajectory the longer the initial phase of GALI$_2
\propto t^{-1}$ lasts, and the smaller is the final non-zero value to
which the index tends.

Let us now perform a more global study of the dynamics of the
H\'{e}non-Heiles system. First, we consider orbits whose initial
conditions lie on the $p_y=0$ line of the PSS of
Fig.~\ref{2D_Ham_HH1}(a). In particular, we use 7000 equally spaced
initial conditions on this line and compute their GALI$_2$ values,
using for each of them the same set of initial (random and
orthonormal) deviation vectors. In Fig.~\ref{nearSPO}(b) we plot the
GALI$_2$ values at $t=10^5$ as a function of $y$.  The regions where
GALI$_2$ has large values ($\gtrsim 10^{-1}$) correspond to regular
motion on 2-dimensional tori. Regions where GALI$_2$ has very small
values ($\lesssim 10^{-12}$) correspond to chaotic or unstable
periodic orbits, while domains with intermediate values
($10^{-4}\gtrsim \mbox{GALI}_2 \gtrsim 10^{-12}$), correspond to
sticky, chaotic orbits. We also distinguish narrow regions where
GALI$_2$ decreases abruptly to values $10^{-1}\gtrsim \mbox{GALI}_2
\gtrsim 10^{-4}$. These correspond to domains of regular motion around
the main stable periodic orbits of the system, as e.g. in the vicinity
of $y \approx 0.3$ which corresponds to the stable periodic orbit in
the center of the main island of stability in the PSS of
Fig.~\ref{2D_Ham_HH1}(a). This behavior appears because GALI$_2$ at
stable periodic orbits decays following a $t^{-1}$ power law and
reaches values smaller than the ones obtained for the neighboring
regular orbits, where GALI$_2$ tends to constant non-zero values, as
we have seen in Fig.~\ref{nearSPO}(a).

This information can be directly used to identify the location of
stable periodic orbits. In Fig.~\ref{nearSPO}(c) we show a color plot
of the parametric space $(H_2,p_y)$ of the H\'{e}non-Heiles system
(\ref{2DHH}). Each point corresponds to an initial condition and is
colored according to its $\log (\mbox{GALI}_2)$ value computed at
$t=10^4$. Chaotic orbits are characterized by very small GALI$_2$
values and are located in the purple colored domains. The deep orange
colored ``strip'' corresponds to the vicinity of a family of stable
periodic orbits (this family is denoted by a white curve) for which
GALI$_2$ attains smaller (but not too small) values with respect to
the surrounding light orange colored region, where regular motion on
2-dimensional tori takes place. We note that, as $H_2$ increases, the
periodic orbit changes its stability and becomes unstable for $H_2
\gtrsim 0.146$. The point $H_2=0.125$, $p_y=0.14979$, denoted by a
black filled circle in Fig.~\ref{nearSPO}(c), corresponds to the
stable periodic orbit of Fig.~\ref{2D_Ham_HH1}(b).

\begin{figure}[!ht]
\centering
  \includegraphics[width=5.15cm]{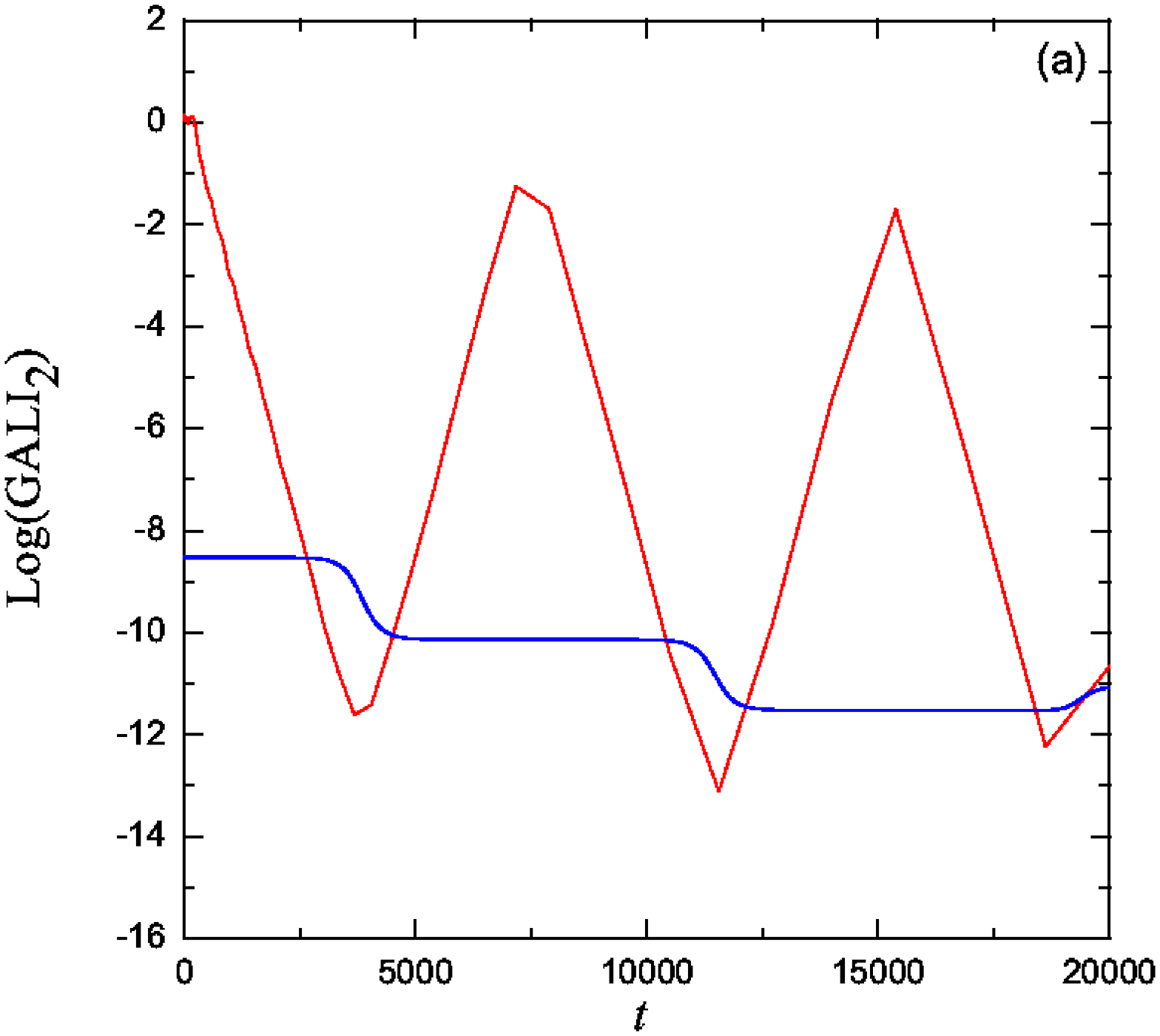}
  \includegraphics[width=5.5cm]{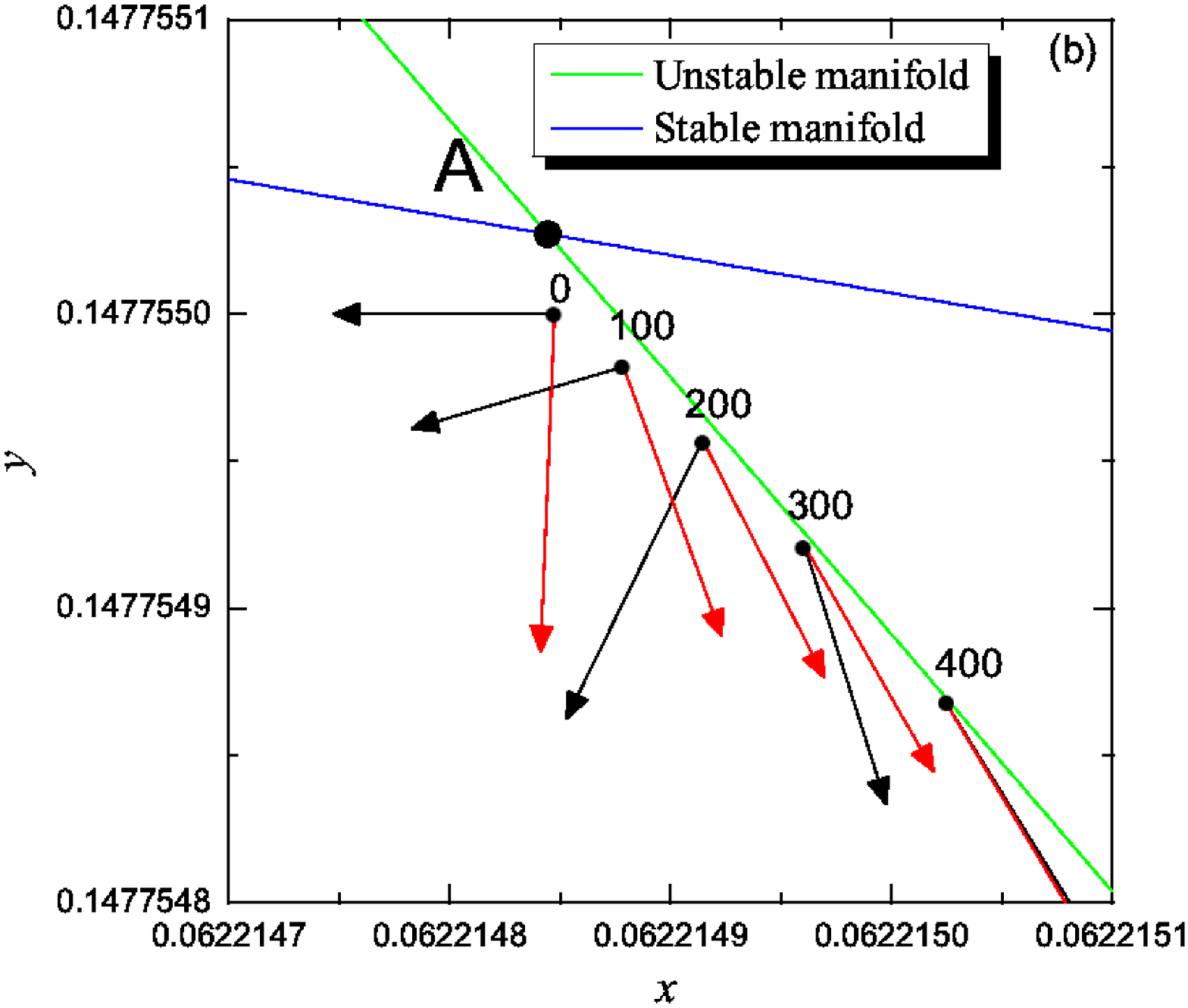}
  \includegraphics[width=5.6cm]{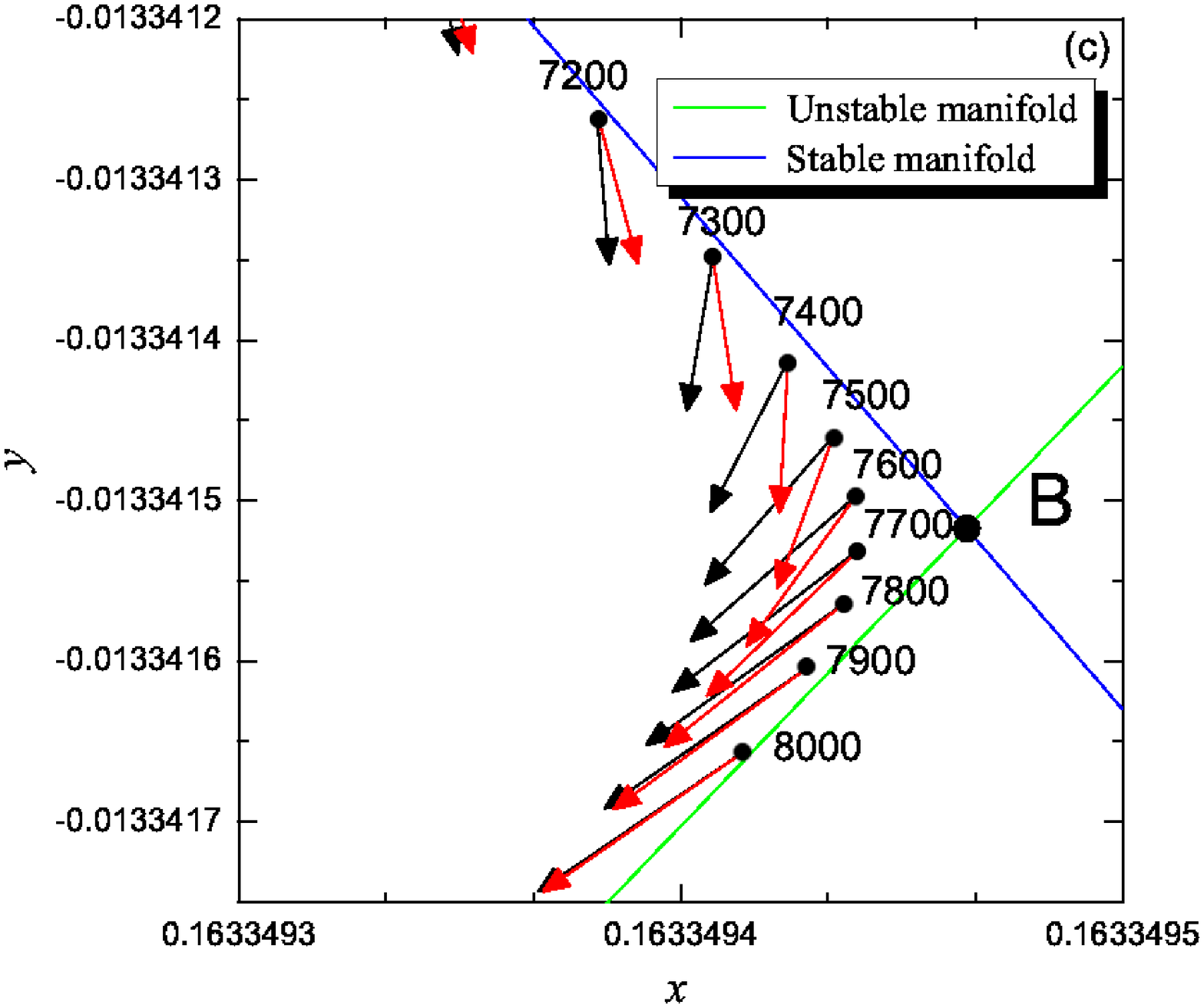}
  \caption{(a) The time evolution of GALI$_2$ of a chaotic orbit of
    the 2D map (\ref{2DHenMap}), with initial condition close to the
    unstable periodic orbit discussed in Sect.~\ref{sect:2DHM}. The
    blue curve shows the $y$ coordinate of the orbit in arbitrary
    units. Consequents of this orbit and of two unit deviation vectors
    from it in the neighborhood of points A and B of the unstable
    periodic orbit of Fig.~\ref{2DHM:1}(a), are respectively plotted
    in (b) and (c). In (b) and (c) the stable and unstable manifolds
    of points A and B are respectively plotted, while the points of
    the chaotic orbit are labeled according to their iteration
    number.}
\label{neighUPO}
\end{figure}

The GALIs of chaotic orbits in the vicinity of unstable periodic
orbits can exhibit a remarkable oscillatory behavior. Such an example
is shown in Fig.~\ref{neighUPO} for a chaotic orbit of the 2D map
(\ref{2DHenMap}) with initial condition
$(x,y)=(0.06221484498946357,0.14775502681732178)$ (point denoted by
`0' in Fig.~\ref{neighUPO} (b)), which is located very close to the
unstable periodic orbit of period 5 discussed in Sect.~\ref{sect:2DHM}
(point A in Fig.~\ref{2DHM:1}(a) and Fig.~\ref{neighUPO} (b)). In
Fig.~\ref{neighUPO}(a) we see that the GALI$_2$ of this orbit
decreases exponentially, reaching very small values (GALI$_2 \approx
10^{-12}$), since the two initially orthonormal deviation vectors tend
to align (Fig.~\ref{neighUPO}(b)) due to the chaotic nature of the
orbit.

The evolution of these vectors is strongly influenced by the stable
and unstable manifolds of the nearby unstable periodic orbit. In
particular, as the chaotic orbit moves away from point A along a
direction parallel to the unstable manifold (green curve in
Fig.~\ref{neighUPO}(b)), both deviation vectors are stretched in this
direction, and shrunk in the direction of the stable manifold (blue
curve in Fig.~\ref{neighUPO}(b)). So, after a few hundreds of
iterations, while the orbit remains in the proximity of point A (note
the tiny intervals in both axes of Fig.~\ref{neighUPO}(b)), the
evolved unit deviation vectors become almost identical, and
consequently GALI$_2$ decreases significantly.

Nevertheless, the angle between the two vectors does not vanish, and
starts to grow again when the orbit approaches point B of
Fig.~\ref{neighUPO}(c), which is the next consequent of the unstable
periodic orbit (see also Fig.~\ref{2DHM:1}(a)). The chaotic orbit
approaches point B moving parallel to the stable manifold of point B
(blue curve in Fig.~\ref{neighUPO}(c)). Now the deviation vectors
start to shrink along this manifold, while they expand along the
direction of the unstable manifold of point B (green curve in
Fig.~\ref{neighUPO}(c)). This leads to a significant increase of the
angle between the two unit vectors, as we see in
Fig.~\ref{neighUPO}(c), and consequently to an increase of the
GALI$_2$ values (Fig.~\ref{neighUPO}(a)).

This oscillatory behavior is repeated as the chaotic orbit visits all
consequents of the unstable periodic orbit, and is clearly seen in
Fig.~\ref{neighUPO}(a) where the $y$ coordinate of the chaotic orbit
is plotted in arbitrary units (blue curve) together with the GALI$_2$
values. The horizontal segments of this curve correspond to the time
intervals that the orbit spends close to the fixed points of the
unstable periodic orbit. During the first part of these intervals the
chaotic orbit approaches a fixed point, the two deviation vectors
become different and GALI$_2$ increases, while afterwards, the chaotic
orbit moves away from the fixed point, whence the two deviation
vectors tend to align, and GALI$_2$ decreases. GALI$_2$ reaches its
lowest values during the short transition intervals between the
neighborhoods of two successive points of the unstable periodic orbit,
which correspond to the short connecting segments between the plateaus
of the blue curve in Fig.~\ref{neighUPO}(a). These oscillations of
GALI$_2$ can last for quite long time intervals, but eventually the
chaotic orbit will escape from the strong influence of the homoclinic
tangle of the unstable periodic orbit and GALI$_2$ will rapidly tend
to zero. It is worth mentioning that abrupt changes in the values of
SALI (which practically is GALI$_2$) by many orders of magnitude was
also reported in \cite{VOYA08} for chaotic orbits of planetary
systems.

\begin{figure}[!ht]
\centering
  \includegraphics[width=5.cm]{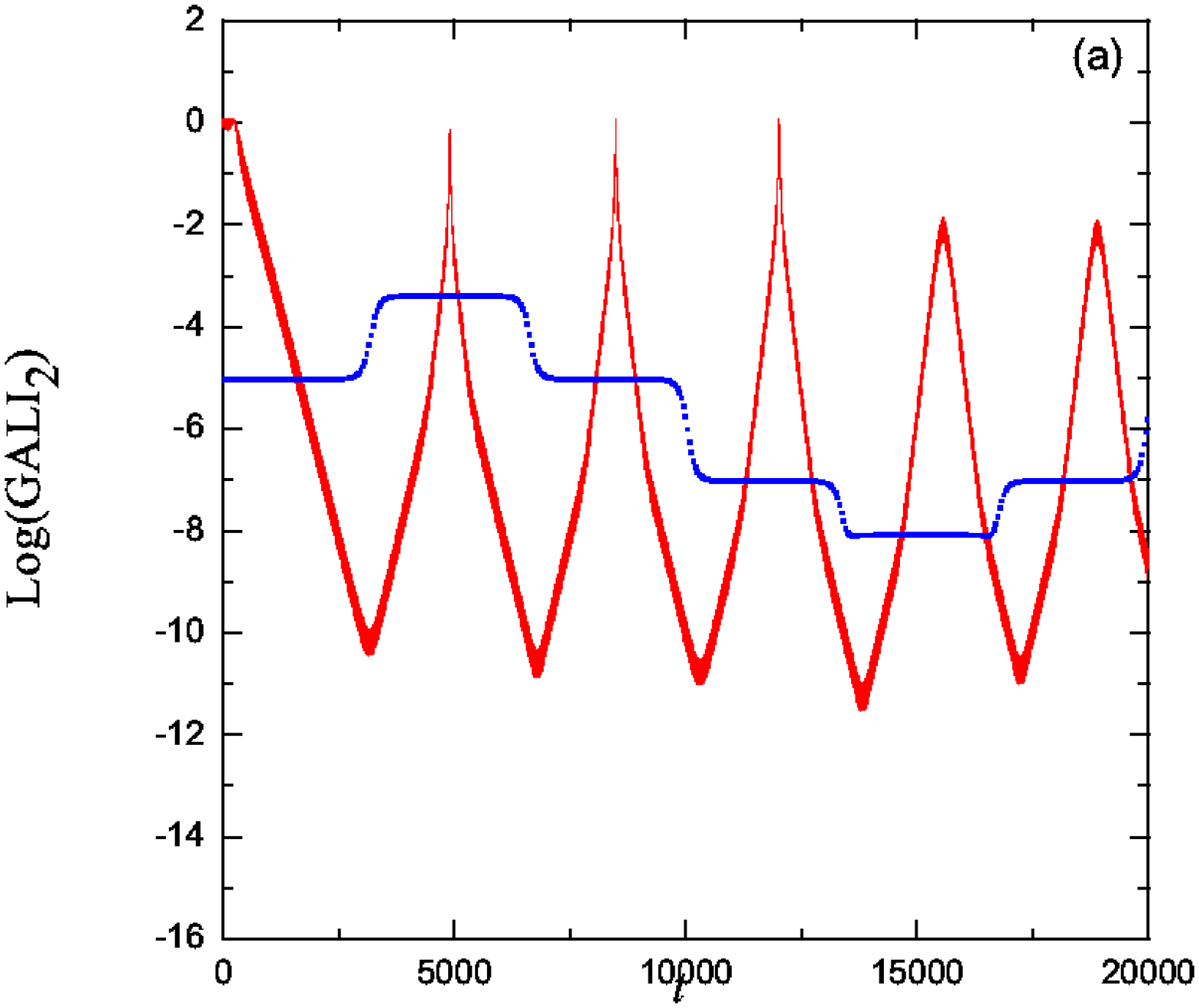}
  \includegraphics[width=5.cm]{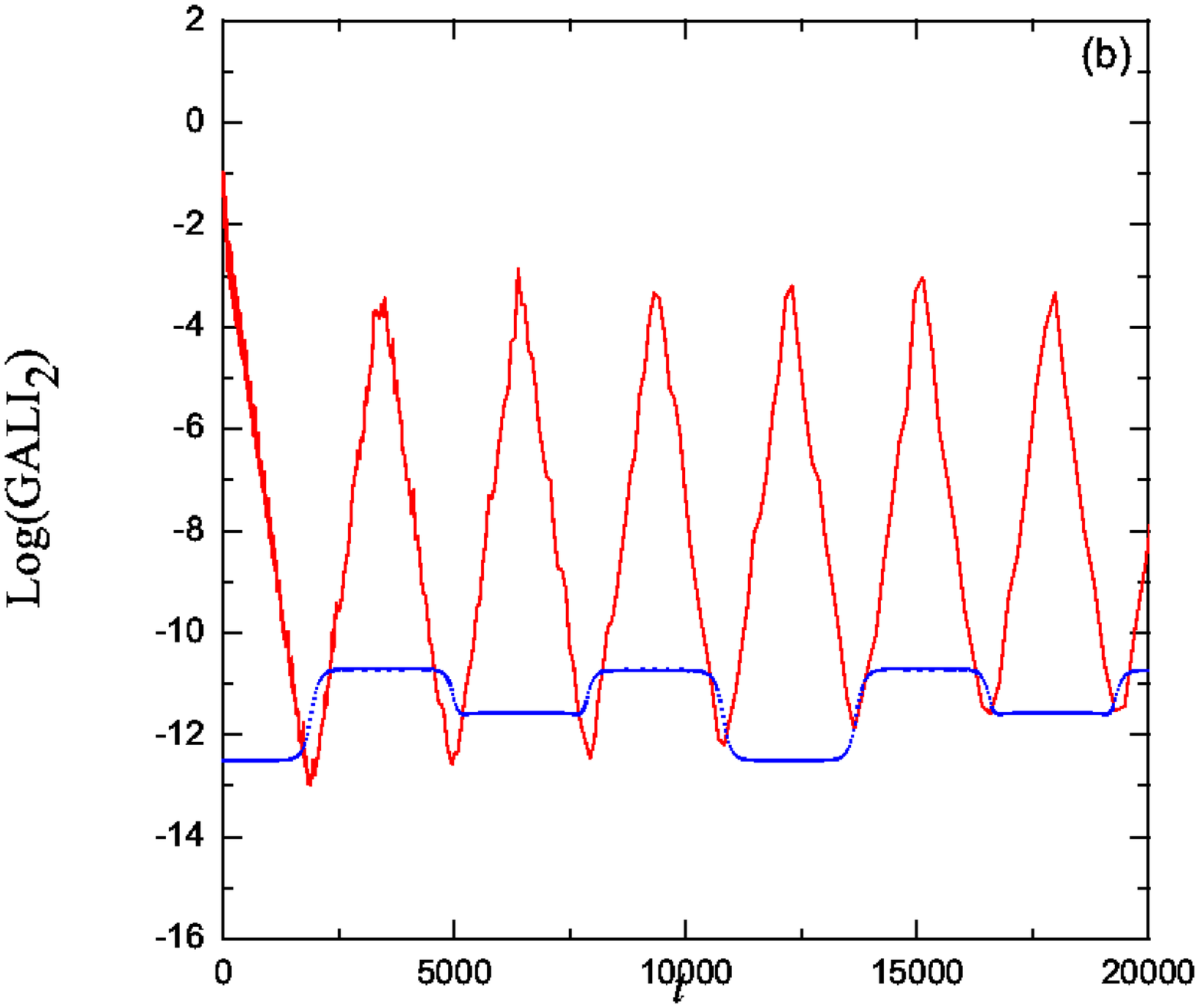}
  \caption{Plots similar to Fig.~\ref{neighUPO}(a) for orbits of (a)
    the 2dof H\'{e}non-Heiles system (\ref{2DHH}), and (b) the 3dof
    Hamiltonian system (\ref{3DHam}). Blue curves show in arbitrary
    units the $y$ coordinate of the studied orbits on (a) the PSS
    $x=0$, $p_x \geq 0$ of system (\ref{2DHH}), and (b) the PSS $z=0$,
    $p_z \geq 0$ of system (\ref{3DHam}).}\label{Ham_neighUPO}
\end{figure}

Up to now, we have described in detail these oscillations of GALI$_2$
in the case of the 2D map (\ref{2DHenMap}) because they can be easily
explained, while the deviation vectors themselves can be visualized in
the 2-dimensional phase space of the map. Interestingly, this
remarkable behavior occurs in higher dimensional systems as well. In
Fig.~\ref{Ham_neighUPO} we show two such examples. In particular, we
consider a chaotic orbit of the 2dof Hamiltonian system (\ref{2DHH}),
whose initial condition is located close to an unstable periodic orbit
of period 7 with initial condition $(x,y,p_x,p_y) \approx
(0,0.1282112414, 0.4847338571,0)$ (Fig.~\ref{Ham_neighUPO}(a)), and an
orbit of the 3dof system (\ref{3DHam}) whose initial condition is near
the $S_1U_1$ periodic orbit presented in Sect.~\ref{sect:3dof}
(Fig.~\ref{Ham_neighUPO}(b)). In both panels of
Fig.~\ref{Ham_neighUPO} we observe an oscillatory behavior of
GALI$_2$, similar to the one shown in Fig.~\ref{neighUPO}(a). We also
point out that in both cases all other GALIs show similar oscillatory
behaviors.

\section{Connection Between the Dynamics of Flows and Maps}
\label{sect:perp_dev_vec}

In Sect.~\ref{sec:stability} we discussed the dynamical equivalence
between $N$dof Hamiltonian systems and $2(N-1)$D maps, as the latter
can be interpreted as appropriate PSSs of the former. We have also
seen that GALIs behave differently for flows and maps. In particular,
as was shown in Sect.~\ref{sec:G_PO}, they remain constant for stable
periodic orbits of maps (see Eq.~(\ref{eq:GALI_PO_maps})) and decrease
to zero for flows, according to Eq.~(\ref{eq:GALI_PO_Ham}).

The fact that maps can be considered as PSS of flows, however, is the
key to understanding this difference. So, computing the restriction of
the GALIs on the PSS of a Hamiltonian system, or more generally on
spaces perpendicular to the flow, should lead to behaviors of the
indices similar to the ones obtained for maps. Actually this approach
has already been successfully applied to other chaos indicators
related to the evolution of deviation vectors \cite{FLFF_02,B_05}, by
only considering the components of these vectors which are orthogonal
to the flow.

\begin{figure}[!ht]
\centering
  \includegraphics[width=5.cm]{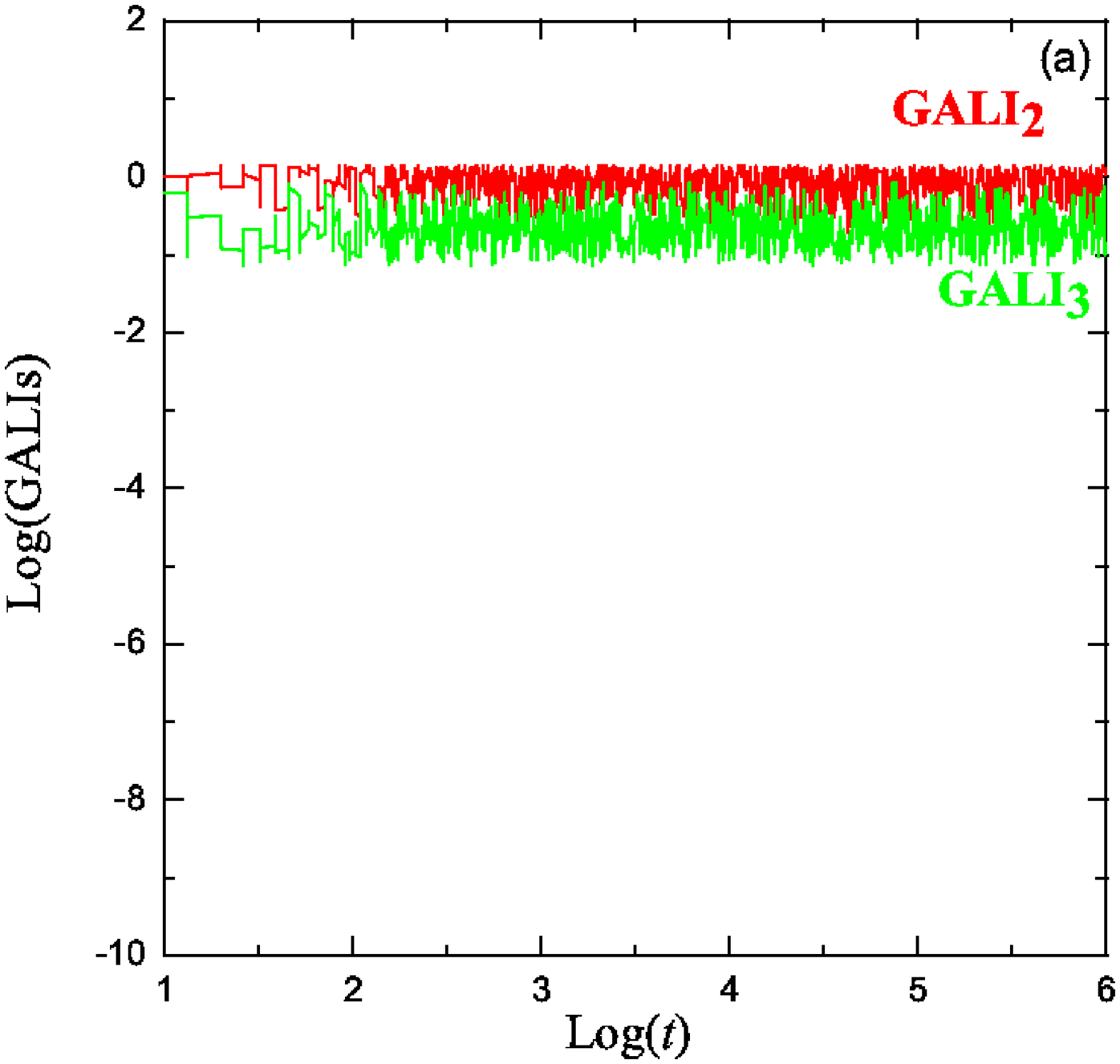}
  \includegraphics[width=5.cm]{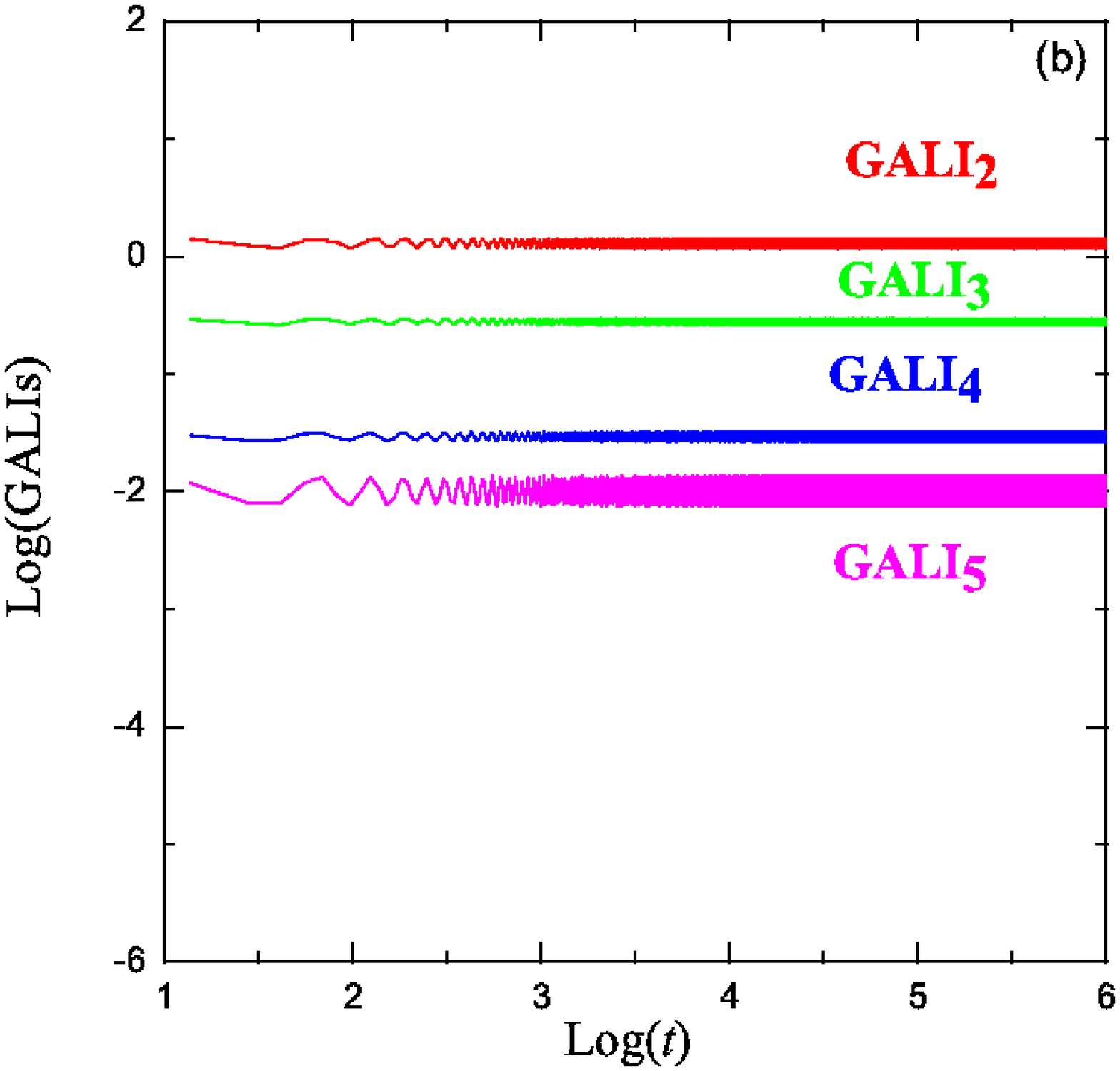}
  \caption{The time evolution of (a) GALI$_k$, $k=2,3$ for the stable
    periodic orbit of the 2dof system (\ref{2DHH}) presented in
    Fig.~\ref{2D_Ham_HH1}(b), and (b) GALI$_k$, $k=2,3,4,5$ for the
    stable periodic orbit of the 3dof system (\ref{3DHam}) presented
    in Fig.~\ref{SPOUPO}(a), when the orthogonal to the flow
    components of the deviation vectors are used.}\label{perpen:1}
\end{figure}

Using deviation vectors orthogonal to the flow, we indeed obtain the
same GALI behavior for stable periodic orbits of flows and maps. Now,
for stable periodic orbits of flows the GALIs of these vectors remain
constant, as we see from Figs.~\ref{perpen:1}(a) and (b) where these
GALIs are plotted for the stable periodic orbits of
Figs.~\ref{2D_Ham_HH1}(b) and \ref{SPOUPO}(a), respectively. These
behaviors differ, however, from the ones shown in
Figs.~\ref{2D_Ham_HH1}(b) and \ref{SPOUPO}(a) where the GALIs of the
usual deviation vectors were computed. We note that when vectors
orthogonal to the flow are used, GALI$_{2N}$ of an $N$dof Hamiltonian
system is by definition equal to zero, because the $2N$ projected
vectors are linearly dependent on a $(2N-1)$-dimensional space. For
this reason GALI$_4$ and GALI$_6$ are not displayed in
Figs.~\ref{perpen:1}(a) and (b) respectively.

\section{Summary}
\label{sec:conclusions}

In this paper, we have explored in more detail the properties of the
GALI method by using it to study the local dynamics of periodic
solutions of conservative dynamical systems. To this end, we have: a)
theoretically predicted and numerically verified the behavior of the
method for periodic orbits, b) summarized the expected behaviors of
the indices and c) clarified the connection between the behavior of
GALIs for dynamical systems of continuous (Hamiltonian flows) and
discrete (symplectic maps) time.

More specifically, we showed that for stable periodic orbits, GALIs
tend to zero following particular power laws for Hamiltonian flows
(Eq.~(\ref{eq:GALI_PO_Ham})), while they fluctuate around non-zero
values for symplectic maps (Eq.~(\ref{eq:GALI_PO_maps})). In addition,
the GALIs of unstable periodic orbits tend exponentially to zero, both
for flows and maps (Eq.~(\ref{eq:GALI_chaos})).

Finally, we examined the usefulness of the indices in helping us
better understand the dynamics in the vicinity of periodic solutions
of such systems. We explained how, the fact that GALIs attain larger
values \textit{near} stable periodic orbits than \textit{on} the
periodic orbits themselves, can be used to identify the location of
these orbits. We also observed a remarkable oscillatory behavior of
the GALIs associated with the dynamics close to unstable periodic
orbits and explained it in terms of the stable and unstable manifolds
of the periodic orbit, showing how the influence of these manifolds
can lead to large variations of the GALI values by many orders of
magnitude.

\nonumsection{Acknowledgments}

The authors thank T.~Bountis for many valuable suggestions and
comments on the content of the manuscript. Ch.~S.~would like to thank
A.~Celletti and A.~Ponno for useful discussions and T.~M.~the Max
Planck Institute for the Physics of Complex Systems in Dresden,
Germany, for its hospitality during his visit in May - June 2009,
where a significant part of this work was performed. This work was
partly supported by the European research project ``Complex Matter'',
funded by the GSRT of the Ministry Education of Greece under the
ERA-Network Complexity Program. Ch.~A.~was also supported by the PAI
2007-2011``NOSY-Nonlinear systems, stochastic processes and
statistical mechanics'' (FD9024CU1341) contract of ULB.


\end{document}